# Two-electron-spin ratchets as a platform for microwave-free dynamic nuclear polarization of arbitrary material targets

Pablo R. Zangara[1], Jacob Henshaw[1], Daniela Pagliero[1], Ashok Ajoy[3], Jeffrey A. Reimer[4], Alexander Pines[3], and Carlos A. Meriles[1,2,†]

[1]Dept. of Physics, CUNY-City College of New York, New York, NY 10031, USA.

[2]CUNY-Graduate Center, New York, NY 10016, USA.

[3]Department of Chemistry, University of California Berkeley, and Materials Science Division Lawrence Berkeley National Laboratory, Berkeley, California 94720, USA.

[4]Department of Chemical and Biomolecular Engineering, and Materials Science Division Lawrence Berkeley National Laboratory University of California, Berkeley, California 94720, USA.

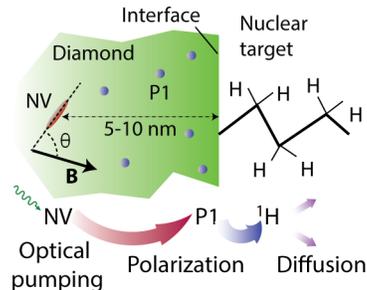

**ABSTRACT**: Optically-pumped color centers in semiconductor powders can potentially induce high levels of nuclear spin polarization in surrounding solids or fluids at or near ambient conditions, but complications stemming from the random orientation of the particles and the presence of unpolarized paramagnetic defects hinder the flow of polarization beyond the defect's host material. Here, we theoretically study the spin dynamics of interacting nitrogen-vacancy (NV) and substitutional nitrogen (P1) centers in diamond to show that outside protons spin-polarize efficiently upon a magnetic field sweep across the NV–P1 level anti-crossing. The process can be interpreted in terms of an NV–P1 spin ratchet, whose handedness —and hence the sign of the resulting nuclear polarization — depends on the relative timing of the optical excitation pulse. Further, we find that the polarization transfer mechanism is robust to NV misalignment relative to the external magnetic field, and efficient over a broad range of electron-electron and electron-nuclear spin couplings, even if proxy spins feature short coherence or spin-lattice relaxation times. Therefore, these results pave the route towards the dynamic nuclear polarization of arbitrary spin targets brought in proximity with a diamond powder under ambient conditions.

The creation of athermal nuclear spin states — i.e., whose absolute polarization is above that defined by Boltzmann statistics — is presently the center of a broad effort encompassing physics, chemistry, and materials science[1]. 'Solid-effect'-based schemes at high magnetic fields — involving microwave (mw) excitation of a radical embedded in a solid matrix — are presently prevalent, but their technical complexity (and corresponding cost) is driving a multi-pronged search for alternative pathways. Among the most promising routes is the use of color centers in insulators including, e.g., the negatively charged nitrogen-vacancy[2-5] (NV) and other nitrogen-related[6] centers in diamond, or the neutral di-vacancy center in silicon carbide[7]. Unlike other semiconductor-hosted paramagnetic defects (already exploited for dynamic nuclear polarization[8,9]), optical excitation spin-polarizes these color centers almost completely, even under ambient conditions. Therefore, rather than generating a *relative* polarization enhancement (crudely proportional to the operating magnetic field **B** and ratio $\mu_e/\mu_n$ between the electron and nuclear magnetic moments), optically-pumped color centers are capable of inducing high *absolute* nuclear spin Zeeman order at low fields[2,10,11]. Further, because of their comparatively long spin-lattice and coherence lifetimes, these spin-active color centers are amenable to electron/nuclear manipulation schemes difficult to implement in optically pumped organic molecules (where spin polarization builds up from short-lived excited triplets[12]).

While the larger surface-to-volume ratio makes the use of powdered semiconductors (as opposed to bulk crystals) better tailored to polarizing a surrounding matrix, the unavoidable misalignment between the applied magnetic field and the color center symmetry axis substantially complicates the transfer of magnetization[13]. In prior work, we demonstrated that $^{13}$C spins in NV-hosting diamond particles can be efficiently polarized through the combined use of continuous optical excitation and mw frequency sweeps[14,15]. To reach nuclear spins outside the diamond lattice, however, polarization must spin diffuse from carbons adjacent to the NV, a relatively slow process that can be hampered by the presence of other unpolarized paramagnetic defects.

Here we theoretically explore an alternate, mw-free approach where shallow paramagnetic defects operate as proxy spins to mediate the transfer of polarization from a source color center deeper in the host lattice to outside nuclei. For concreteness, we consider the case of NV centers and substitutional nitrogen in diamond — the so-called 'P1 center' — but the ideas we lay out can be naturally extended to other spin systems. Our approach exploits the 'cross effect', where nuclear spins polarize thanks to the slight energy mismatch between the source and proxy spins[16-18]. Unlike traditional, mw-mediated implementations, however, here we show that protons proximal to P1 centers can be polarized efficiently through protocols articulating pulsed optical excitation and low magnetic field sweeps. Remarkably, we attain additive dynamic nuclear polarization during both the 'low-to-high' and 'high-to-low' segments of the field sweep cycle, with the sign of the nuclear polarization depending on the relative timing of the optical excitation pulse. Further, we show that proxy spins featuring short coherence or spin-lattice relaxation times can efficiently contribute to the process, and that the technique is robust over a broad set of inter-spin couplings and relative orientations of the NV axis and **B** field. For completeness, we note that related studies involving optically-initialized color centers and proxy spins have been discussed in the context of nanoscale quantum sensing[19-21].

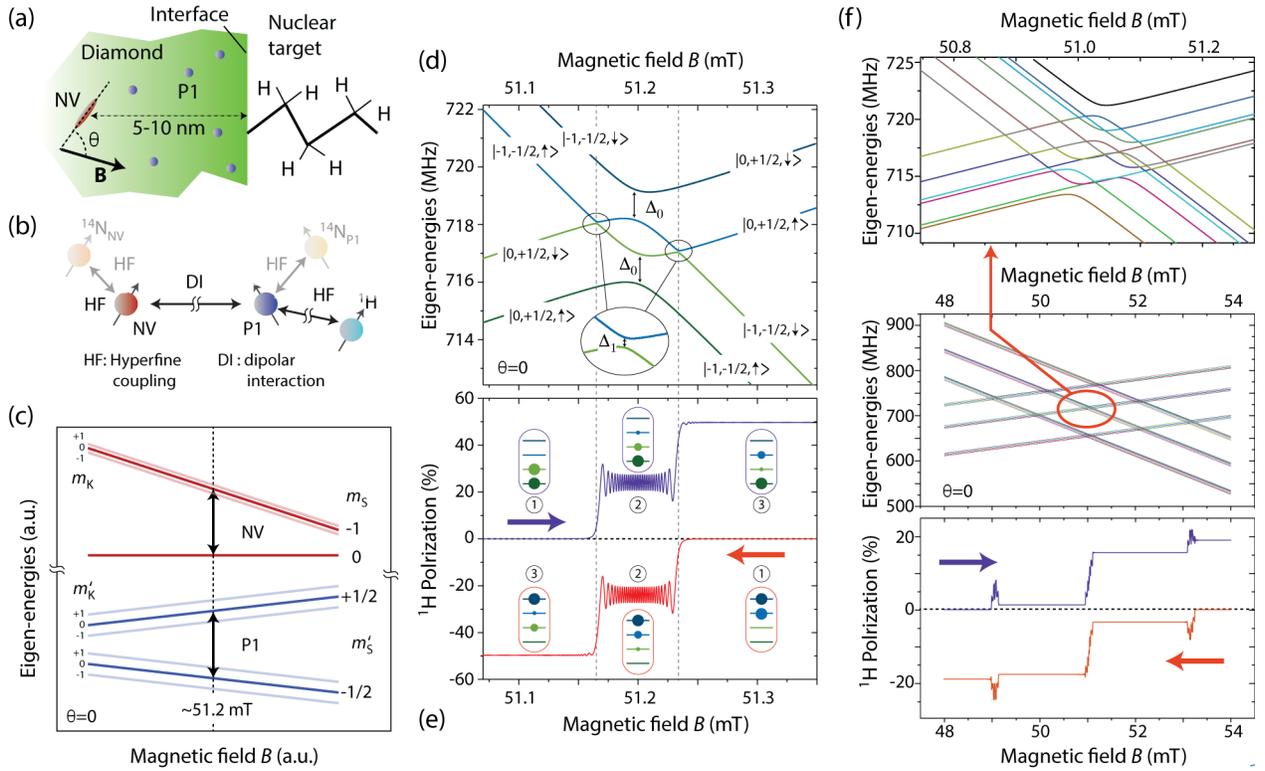

**Figure 1. Proxy-spin-mediated polarization transfer to weakly-coupled proton spins.** (a) Typically, the NV and target outer nuclear spins (protons in this case) are separated by at least 5-10 nm, thus leading to comparatively weaker couplings with protons on the surface; more proximal P1 centers (and other spin-1/2 surface defects) can thus serve as proxies to mediate the polarization transfer. (b) Electron/nuclear spin model; unless explicitly noted, we ignore the NV and P1 hyperfine interactions with their $^{14}$N hosts (faded spins). (c) Schematic energy diagram for the NV and P1 centers; near 51 mT, the NV $|m_S = 0\rangle \leftrightarrow |m_S = -1\rangle$ and P1 $|m'_S = +1/2\rangle \leftrightarrow |m'_S = -1/2\rangle$ transitions are nearly matched. (d) Level anti-crossing near 51 mT for the combined NV–P1–$^1$H spin system without considering the hyperfine couplings with the $^{14}$N spins. (e) Calculated $^1$H spin polarization as we sweep the magnetic field from left to right or from right to left (purple and red traces, respectively). Inserts indicate relative populations using the color code in (d). (f) Same as in (d,e) but taking into account the NV and P1 hyperfine couplings with their respective $^{14}$N hosts; the upper plot is a blown out view of the energy level structure within the circled region. Throughout these calculations, we assume the NV axis and external magnetic field are parallel ($\theta = 0$), the NV–P1 coupling is 500 kHz, the P1–$^1$H coupling is 200 kHz, and the field sweep rate is 0.26 mT/ms; we also assume optical excitation pumps the NV spin to 100% and all other spin species are unpolarized.

Fig. 1a sketches the starting nanoscale geometry: NV centers — in general, arbitrarily oriented relative to **B** — coexist with more abundant P1 centers and other shallow paramagnetic defects (e.g., dangling bonds at the diamond surface, not shown). Band bending effects — prone to remove the excess electron from the negatively charged NV — impose a minimum distance to the surface[22] (at best of order 5-10 nm), meaning that the NV interaction with outside nuclear spins (protons in the case considered herein) is comparatively weak. Paramagnetic centers in the region separating the NV from the nuclear target are ideally positioned to mediate the transfer of polarization, but their faster transverse and spin-lattice relaxation as well as the broad distribution of coupling strengths between the source, proxy, and target spins poses a number of difficulties seemingly difficult to overcome.

To quantitatively model our polarization scheme, we consider the spin cluster in Fig. 1b comprising an NV coupled to a P1 center (a spin-1/2 defect); the latter also interacts with a proton spin on the diamond surface via a dipolar-type hyperfine coupling. Both the NV and P1 electron spins are hyperfine coupled to the nuclear spins of their respective nitrogen hosts (typically taking the form of $^{14}$N isotopes). We later show, however, these interactions do not significantly impact the polarization flow to the proton, and can thus be ignored (see Hamiltonian formulation in the Supporting Information, Section S.I). For clarity, we first consider the energy diagram for the NV-P1 pair in the case where the NV symmetry axis and magnetic field are collinear ($\theta = 0$). Near 51 mT, the P1 Zeeman splitting matches the energy separation corresponding to the $|m_S = 0\rangle \leftrightarrow |m_S = -1\rangle$ transition of the NV, thus leading to cross-relaxation between both electron spins[23]. Slightly above and below the matching field, energy conservation in the electron-electron spin transition can be regained with the assistance of a coupled nuclear spin, which flips in one direction or the other depending on the sign of the difference between the NV and P1 splittings. Since continuous optical illumination spin pumps the NV into $|m_S = 0\rangle$, cross relaxation leads to P1-assisted dynamic nuclear polarization, whose sign — alternating from positive to negative — depends on the chosen magnetic field[17,18].

The route we pursue herein starts with the use of an optical pulse to initialize the NV spin into $|m_S = 0\rangle$, followed by a gradual change of the magnetic field amplitude $B$ across the range where electron/nuclear cross relaxation takes place. To understand the ensuing spin dynamics, we first plot the energy diagram for the NV-P1-$^1$H set as a function of $B$ in the simplified case where the NV and P1 hyperfine interactions with the nuclear spins of their respective nitrogen hosts is zero (Fig. 1d). We find a series of avoided crossings with energy gaps strongly dependent on the particular pair of eigenstates. In our calculations, we choose the field sweep rate so as to make the Landau Zener (LZ) dynamics partially non-adiabatic at the narrower gaps between the two inner branches (featuring opposite nuclear spin numbers), i.e., we set $\beta \equiv dB/dt$ so that $p_1 > p_0 \sim 0$, where $p_i \sim exp\{-\pi \Delta_i^2 / (4|\gamma_e|\beta)\}$, $i = 0, 1$ is the Landau-Zener probability of bifurcation, $\gamma_e$ is the electron gyromagnetic ratio, and $\Delta_1$ ($\Delta_0$) denotes the narrower (wider) energy gap (Fig. 1d, see analytical estimates for $\Delta_1$ and $\Delta_0$ in the Supporting Information, Section S.II). Therefore, assuming no initial $^1$H magnetization (stage (1) in Fig. 1e), sweep-induced population exchange during the narrower LZ crossings leads to the generation of net nuclear spin polarization (stages (2) and (3)), as confirmed by direct numerical computations both for low-to-high and high-to-low field sweeps (upper and lower plots in Fig. 1e). Note that proton polarization of the opposite sign builds up in one case or the other; this behavior is somewhat reminiscent of that experimentally observed for carbons in NV-hosting diamonds simultaneously subjected to mw frequency sweeps and continuous optical illumination[14]. We later show, however, it is the relative timing of the illumination rather than the sweep direction what defines the sign of the resulting nuclear polarization.

An accurate description of the system spin dynamics must take into account the NV and P1 hyperfine couplings with their nuclear spin hosts. Assuming $^{14}$N isotopes in both cases — featuring spin numbers $K, K' = 1$ — each level in the energy diagram of Fig. 1e splits into nine distinct branches, corresponding to different combinations of the quantum projections $m_K, m_K' = 0, \pm 1$. Given the dominant character of the P1–$^{14}$N hyperfine coupling (of order ~100 MHz), the diagram shows well-resolved eigen-energy sets (center plot in Fig. 1f), but the high multiplicity leads to subtle structures at the avoided crossings (upper insert in Fig. 1f). Despite this complexity, the system response upon a field sweep retains the main traits found in the simpler case (i.e., Fig. 1d), namely, the proton spin polarizes efficiently with a sign dependent on the sweep direction (lower graph in Fig. 1f). Similar to Fig. 1e, we attain near-optimal levels of proton polarization, though quantum interferences during the LZ crossings make the exact value a sensitive function of the sweep rate (see Supporting Information, Section S.III). To gain physical intuition (and speed up computations), we henceforth ignore the hyperfine coupling with the nitrogen hosts with the understanding that these contributions may only slightly alter some of the numerical values we derive, without fundamentally modifying the underlying transfer processes.

Since the spin dynamics is insensitive to the exact start and end magnetic field values, the results in Fig. 1 indicate that P1-assisted DNP can be made robust to field heterogeneities (and, as we show later, to spin coupling dispersion and NV orientation disorder). On the other hand, the slow sweep rate required for optimal efficiency (~0.3 mT/ms) is at odds with the relatively short coherence and spin-lattice relaxation times of both electron spins near the diamond surface. Further, because unlike mw, a magnetic field must be present at all times, the impact of a full field cycle (including the low-to-high and high-to-low ramps) on the end proton polarization is a priori unclear.

We address these issues in Fig. 2a, where we monitor the nuclear spin evolution as we complete successive field cycles using a tenfold faster sweep rate (3 mT/ms). Somewhat unexpectedly, we find that nuclear polarization adds constructively during both halves of the field cycle, with the sign being determined by the illumination timing rather than the sweep direction: Positive (negative) $^1$H polarization emerges from NV spin initialization at the low field (high-field) extremum of the cycle (Figs. 2a and 2b, respectively). Consistent with this response, we observe negligible DNP for a cycle with optical illumination at both extrema (Fig. 2c), with the imperfect cancellation between the low-to-high and high-to-low halves arising from the slight asymmetry in the initial spin populations at each half period during the first few repeats (see Supporting Information, Sections S.IV).

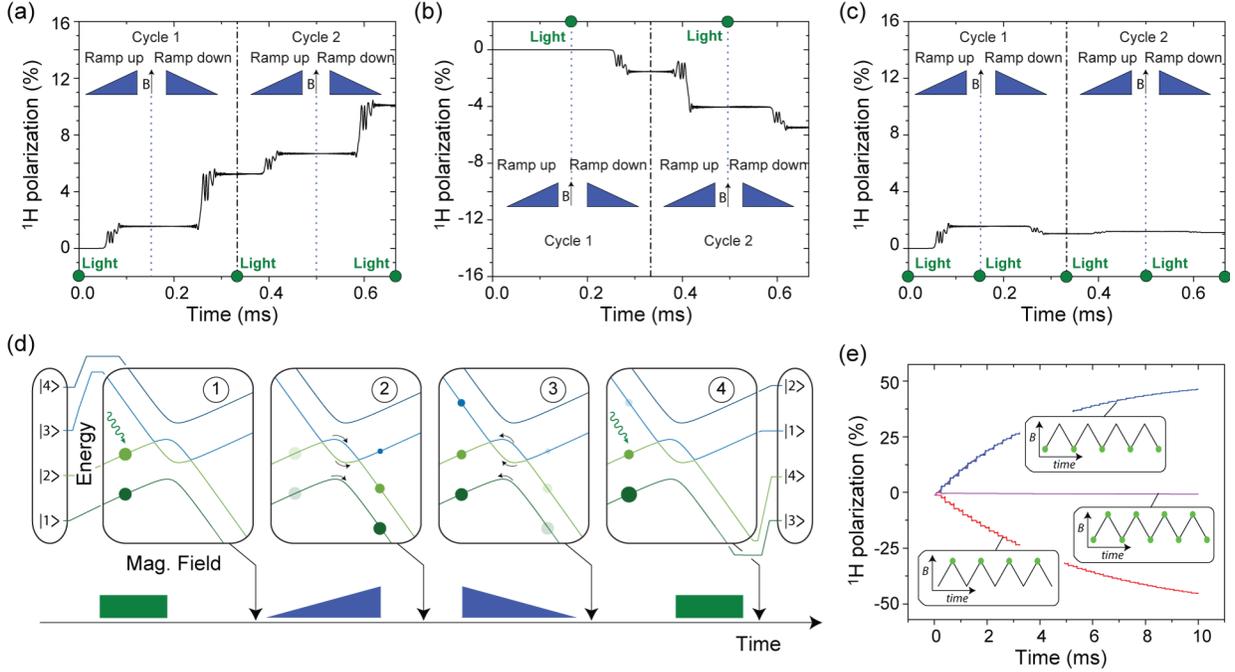

**Figure 2. The two-electron spin ratchet.** (a) Proton dynamic polarization upon two full magnetic field cycles; the field range is centered around the NV–P1–$^1$H level anti-crossing at ~51 mT and NV spin preparation takes place at the low field point. (b) Same as in (a) but assuming optical excitation at the time when the magnetic field is at its maximum. (c) Same as in (a) but for a sequence where NV spin pumping takes place both at the beginning and midpoint of each cycle. (d) Schematics of the proton polarization buildup upon application of the protocol in (a). Green squares (blue triangles) indicate laser pulses (magnetic field ramps). Rounded squares in the upper half reproduce the energy diagram from Fig. 1a. Spin populations are represented by solid circles of variable radius; for clarity, we assume the laser pulse fully projects the NV spin into $m_S = 0$, though only partial spin pumping is required. Here we use the state notation $|1\rangle = |0, +1/2, \uparrow\rangle$, $|2\rangle = |0, +1/2, \downarrow\rangle$, $|3\rangle = |-1, -1/2, \uparrow\rangle$, $|4\rangle = |-1, -1/2, \downarrow\rangle$, with labels representing the NV, P1, and $^1$H quantum projections, respectively. (e) Evolution of the $^1$H spin polarization as a function of time upon multiple repetitions of the optical excitation/field sweep cycles described above. Throughout these calculations, $\theta = 0$, we use $\dot{v}_B = 3$ mT/ms, and the field range is 0.5 mT. Further, the NV–P1 and P1–$^1$H couplings are respectively 500 kHz and 100 kHz, and the NV spin polarization upon optical excitation is 100%; we consider no transverse or longitudinal relaxation processes.

The formal description of the process is not simple, but can be attained with the use of a transfer matrix (TM) model (see Supporting Information, Section S.V). To qualitatively illustrate the underlying dynamics, Fig. 2d follows the evolution of the proton spin polarization throughout a cycle of magnetic field sweeps and optical excitation at the low-field-end of the ramp (the protocol in Fig. 2a); for simplicity, we assume that laser excitation fully spin pumps the NV into $m_S = 0$, though we note that only a partial spin projection is required. In the regime of moderately fast field sweeps where $p_1 \lesssim 1$ and $p_0 \sim 0$, both inner branches nearly exchange their populations during each of the two LZ crossings, while the populations of the outer branches remain unchanged. Correspondingly, only a small (positive) nuclear spin imbalance emerges from the low-to-high field ramp (stage (2) in Fig. 2d). As a simple visual inspection shows, however, this difference virtually doubles if one subsequently forces the system to undergo a reversed, high-to-low field ramp (stage (3) in Fig. 2d). Subsequent optical excitation — acting exclusively on the NV — leaves the nuclear spin polarization unchanged and thus resets the system for a new DNP cycle (stage (4) in Fig. 2d).

Repeated application of the same protocol leads to gradual accumulation of proton polarization in the form of an exponential growth towards a near-optimum value approaching the starting NV spin polarization (Fig. 2d). This one-directional flow of spin polarization upon a time-periodic, zero-mean modulation of the magnetic field is analogous to the directed motion of quantum motors[24], prompting us to interpret the dynamics as that of a two-electron spin ratchet. Note that in the absence of transverse or spin lattice relaxation — the limit assumed so far, see below — the time interval between successive optical pulses can be increased so as to encompass multiple field cycles $l > 1$. The corresponding response — in general, a function of $l$ — remains comparable to that shown in Fig.

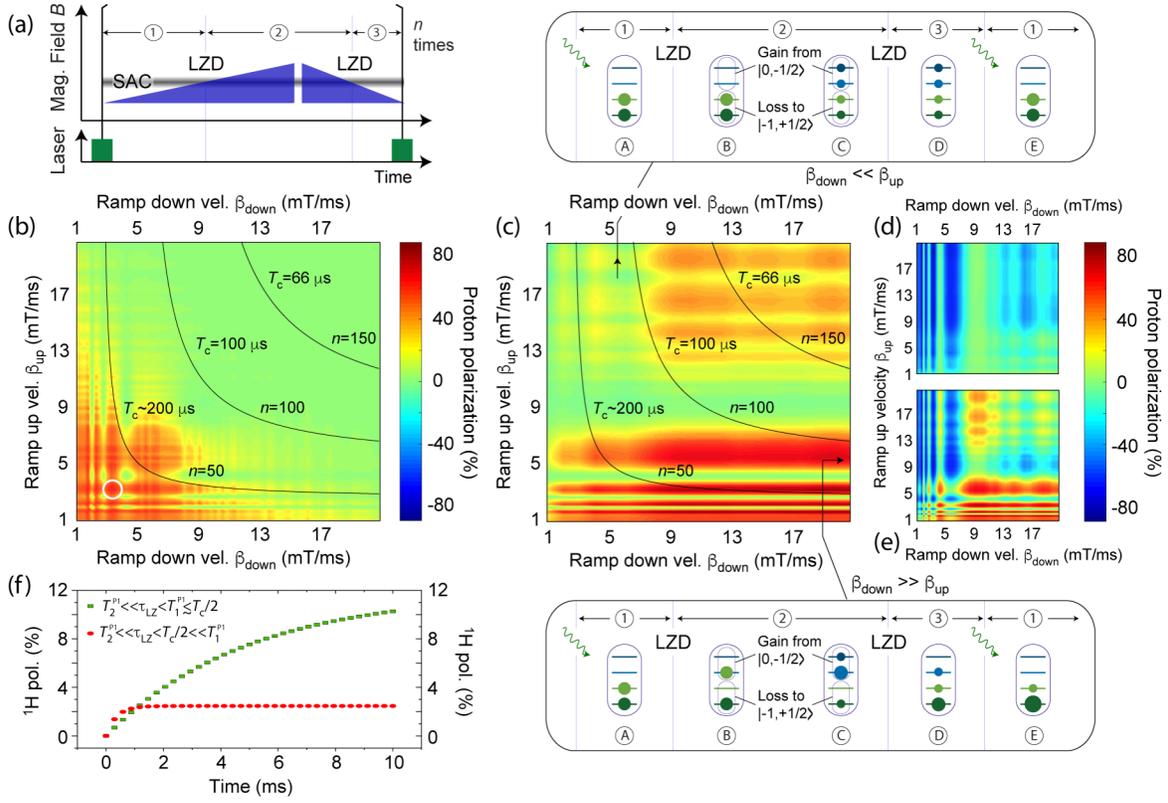

**Figure 3. Proxy-spin-mediated DNP in the presence of spin-lattice relaxation.** (a) Generalized DNP protocol featuring variable low-to-high and high-to-low field sweep velocities. (b) Calculated $^1$H spin polarization as a function of the 'ramp up' and 'ramp down' velocities for a total DNP time of 10 ms in the regime $\tau_{LZ} < T_2^{(P1)} \leq T_c/2 \ll T_1^{(P1)}$; optical excitation takes place at the low-field end of the ramp. Solid lines indicate areas of the plot sharing the same cycle time $T_c$, and hence undergoing the same number $n$ of DNP repeats. (c) Same as in (b) but assuming the P1 spin relaxes before and after traversing the set of avoided crossings so that $\tau_{LZ} < T_2^{(P1)} \leq T_c/2 \sim T_1^{(P1)}$. The upper and lower inserts highlight the impact of P1 spin-lattice relaxation throughout the DNP cycle in (a) for the limit cases where one field ramp is much faster than the other one; for simplicity, we collapse the NV–P1–$^1$H energy diagrams to sets of four horizontal lines, each corresponding to the branch in Fig. 1d with the same color code. (d) Same as in (c) but for optical excitation at the high-field end of the ramp. (e) Same as in (c) but assuming optical excitation both at the low- and high-field ends of the ramp. (f) Proton spin polarization buildup as a function of time upon repeated applications of the DNP protocol when $T_2^{(P1)} \ll \tau_{LZ} < T_c/2 \ll T_1^{(P1)}$ (red ellipses) or when $T_2^{(P1)} \ll \tau_{LZ} < T_c/2 \sim T_1^{(P1)}$ (green rectangles). In both cases, we assume $\beta_{down} = \beta_{up} = 3$ mT/ms and $T_2^{(P1)} = 100$ ns. In all plots, we assume θ = 0, the field range is 0.5 mT, the NV–P1 and P1–$^1$H couplings are respectively 500 kHz and 100 kHz, and the NV spin polarization upon optical excitation is 100%. In (a) and (c), SAC denotes the 'set of avoided crossings' approximately midway within the magnetic field range and LZD indicates Landau-Zener dynamics.

2e, both in terms of the nuclear polarization buildup rate and limit value, though we warn that quantum interference effects between successive passages can make the growth strongly non-monotonic (see Supporting Information, Section S.IV).

Naturally, the field sweep cycle of Fig. 2 can be generalized to the case where the low-to-high and high-to-low field sweep rates (respectively denoted $\beta_{up}$ and $\beta_{down}$) are unequal (Fig. 3a). Fig. 3b shows the calculated proton polarization for periodic illumination at each low-field extremum assuming the total DNP time remains fixed (10 ms). To avoid quantum interferences between successive crossings, we impose the condition $\tau_{LZ} < T_2^{(P1)} < T_c$, where $T_2^{(P1)}$ denotes the P1 transverse coherence lifetime (assumed shorter than $T_2^{(NV)}$), $T_c$ is the polarization cycle period, and $\tau_{LZ} \sim \Delta_0/(2|\gamma_e|\beta)$ is the characteristic Landau-Zener time[25], required for a coherent transfer (see below). The interplay between polarization transfer efficiency and multiple repetitions leads to near optimal proton polarization over a broad set of up/down sweep rate combinations, though optimal DNP is restricted to the lower left corner of the plot, corresponding to DNP cycle times $T_c \gtrsim 0.2$ ms.

We are now in a position to investigate the influence of spin relaxation, which, as we show next, can have a profound effect on the system response. For simplicity, we assume that only the P1 spin relaxes during a DNP cycle (i.e., $T_1^{(P1)} < T_1^{(NV)}$), a condition justified in the present case given the imposed P1 proximity to the diamond surface (and/or spin exchange with other paramagnetic defects in its neighborhood). Fig. 3c shows the result of a calculation where we impose $T_1^{(P1)} = T_c/2$ so as to force

P1 spin relaxation after the first (but before the second) LZ pass in a cycle across the set of avoided crossings (SAC). Unlike Fig. 3b — insensitive to an exchange of the up and down field sweep rates — the presence of a finite $T_1^{(P1)}$ time introduces a strong asymmetry (Fig. 3c). Optimal nuclear polarization builds up for arbitrarily large $\beta_{\text{down}}$ rates so long as $\beta_{\text{up}}$ does not exceed an upper threshold (~7 mT/ms for the present set of couplings), but the converse is not true.

The impact of spin relaxation, though complex, can be formally incorporated in our TM approach (see Supplementary Information, Section VI). A simpler, more intuitive understanding of the underlying dynamics, however, can be gained by considering the evolution of spin populations in the regime where the faster passage in a non-symmetric field cycle is fully non-adiabatic (i.e., when $p_0 \sim p_1 \sim 1$, upper and lower inserts in Fig. 3c). If the first half of the cycle is the faster one, spin-lattice relaxation *before* traversing the set of avoided crossing a second time populates all four energy branches equally, meaning that the ensuing LZ dynamics cannot produce nuclear polarization regardless the sweep rate (upper insert in Fig. 3c). The result is different in the converse regime (i.e., when the second sweep is faster), because P1 spin-lattice relaxation does not degrade the nuclear spin population created during the first pass (lower insert in Fig. 3c, and Supplementary Information, Section S.IV). In other words, positive nuclear spin polarization can be produced in the limit where $\beta_{\text{down}} \gg \beta_{\text{up}}$ but the inverse is not true. Interestingly, spin-lattice relaxation can still induce substantial nuclear polarization even when both sweep rates are comparably fast (upper right corner in the main plot of Fig. 3c), through the underlying dynamics is more complex (Supplemental Information, Sections S.IV through S.VI).

For completeness, we note that NV spin initialization at the high-field extremum of the cycle (i.e., the generalization of the protocol in Fig. 2b) simply produces a reversal in the asymmetry, i.e., negative nuclear polarization emerges for $\beta_{\text{down}} \ll \beta_{\text{up}}$ (Fig. 3d). Likewise, optical excitation both at the low- and high-field extrema can yield net nuclear polarization of one sign or the other whenever $\beta_{\text{down}} \neq \beta_{\text{up}}$ (Fig. 3e). As a corollary, continuous laser illumination (as opposed to synchronous, pulsed optical excitation) should yield efficient DNP provided the two sweep rates are substantially different from each other.

Given the short spin lifetimes typical in near-surface paramagnetic centers, the regime $T_2^{(P1)} < \tau_{\text{LZ}}$ — corresponding to the strongly dissipative limit — deserves special consideration. In this regime, coherent transfer of the spin polarization is not possible and the system dynamics is better described via the Landau-Zener formulas for the case of strong-dephasing [26-28] (see Supplementary Information, Sections V and VI). The impact of fast P1 decoherence is illustrated in Fig. 3f assuming $T_2^{(P1)} = 100$ ns and $\beta_{\text{down}} = \beta_{\text{up}} = 3$ mT/ms (white circle on the plot of Fig. 3b): Under these extreme conditions, $T_2^{(P1)} \ll \tau_{\text{LZ}} \sim 5$ μs and, in the limit $T_c \ll T_1^{(P1)}$, we calculate strongly attenuated proton polarization buildup (red circles in Fig. 3f). However, for shorter P1 spin lifetimes $\tau_{\text{LZ}} < T_1^{(P1)} \lesssim T_c/2$, we find that a substantial fraction of the original DNP efficiency can be regained (i.e., P1 spin lattice relaxation partially remedies fast decoherence, green dots in Fig. 3f).

Since in a realistic setting the spatial separations between the source, proxy, and target spins change randomly, efficient polarization transfer to outside nuclei requires the DNP protocol to be robust to spin coupling heterogeneity. We address this issue in Fig. 4a, where we calculate the proton polarization upon application of the sequence in Fig. 2a as a function of both the NV–P1 and P1–$^1$H couplings. Remarkably, we attain near optimal polarization transfer over a broad set of conditions extending down to NV–P1 (P1–$^1$H) couplings as weak as ~300 kHz (~200 kHz). The latter corresponds to source-proxy (proxy-target) spin separations as large as ~5 nm (~1 nm). These distances are typical in samples that have been engineered to host shallow NVs[29], and hence compatible with proxy-spin-mediated polarization transfer to outside nuclear targets (see lower insert in Fig. 4a).

Finally, we investigate the DNP efficiency as a function of the magnetic field orientation relative to the three-spin set. For our calculations, we choose a reference frame whose *z*-axis points along the NV direction, and where the *xz*-plane matches that defined by the NV and the NV-P1 axes (Fig. 4b). In order to study the LZ dynamics away from $\theta = 0$, we first consider an isolated NV–P1 pair and determine the 'matching' field $B_m$ (where the P1 Zeeman splitting coincides with the NV $|m_S = 0\rangle \leftrightarrow |m_S = -1\rangle$ energy difference) as a function of the polar angle (Fig. 4c). Below $\theta \sim 40$ deg., $B_m$ varies over a moderate range (50-90 mT) thus making it possible to envision polarization transfer over a sizable polar cone of relative crystal orientations with only modest practical means. Recent experimental observations demonstrating efficient P1-mediated carbon polarization in diamond for $\theta$ approaching ~20 deg. are consistent with this notion[17].

Fig. 4d shows a polar representation of the proton polarization as a function of the angular coordinates $\theta$ and $\phi$. We find a mild dependence on the azimuthal angle, but the polar response exhibits sharp minima superimposed to an overall decay. The latter stem from quantum interference between consecutive LZ crossings (within the same sweep) and their angular positions depend on the considered couplings and exact conditions of the sweep (see insert to Fig. 4d). The overall envelope, on the other hand, arises partly from modest NV spin initialization $\eta_{NV}$ at higher matching fields $B_m$ (right plot in Fig. 4c) combined with poor polarization transfer efficiency. We therefore conclude this form of proxy-spin mediated DNP is confined to the solid cone defined by $\theta \sim 20$ deg., which, for the case of a powdered diamond sample, corresponds to limiting the field sweep to the range 51–57 mT. This robustness to field misalignment (or, by the same token, field heterogeneity) is in strong contrast with prior demonstrations of P1-assisted DNP[17,18], where contributions from all crystallites — positive or negative depending on the local field strength and/or relative orientation — average out. This problem is particularly acute for weakly coupled nuclei because they polarize (positively or negatively) only in a close vicinity of the matching field[17,18,30].

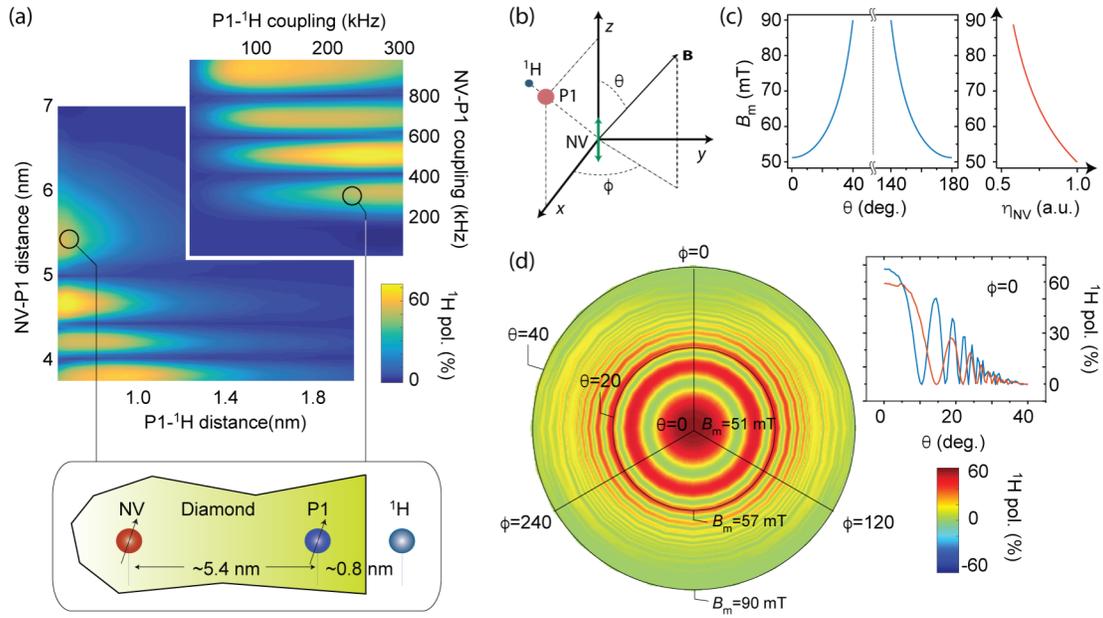

**Figure 4. The impact of spin coupling heterogeneity and orientation disorder.** (a) Proton polarization as a function of the NV–P1 and P1–$^1$H distances. The upper insert shows the same plot as a function of the corresponding coupling strengths. Efficient DNP can be attained for NV–P1 (P1–$^1$H) couplings down to 350 kHz (250 kHz) corresponding to spin distances of ~5.4 nm (~0.8 nm), as sketched in the lower insert. (b) Schematics of the reference frame for the case where the **B** field and the NV symmetry axis are not collinear. For simplicity, we choose the location of the proton along the axis connecting the NV and P1 spins. (c) Matching field $B_m$ as a function of the polar angle $\theta$. The side graph shows the NV optical pumping efficiency $\eta_{NV}$ as a function of $B_m$. (d) Polar representation of the $^1$H polarization as a function of the magnetic field orientation assuming $\beta_{up} = 3.25$ mT/ms and $\beta_{down} = 20$ mT/ms; the plot takes into account the NV spin pumping efficiency $\eta_{NV} \leq 1$ at a given field $B_m$. The blue trace in the upper right insert shows the cross section in the main plot for $\phi = 0$; the red trace provides the same information but for $\beta_{up} = 6$ mT/ms and $\beta_{down} = 10$ mT/ms. In (a) we assume $\theta = 0$, and in (d) we make the NV–P1 (P1–$^1$H) coupling equal to 500 kHz (100 kHz). The total number of DNP cycles at each point is $n = 56$; all other conditions as in Fig. 2a.

Before concluding, we note that in a realistic environment, the NV may simultaneously interact with P1s other than the near-surface proxy we modeled, thus prompting the question as to whether the flow of polarization can be diverted off the target. To investigate this possibility, we consider an extension of the case in Fig. 4a, where the spin cluster is modified to include an additional 'bystander' P1 strongly coupled to the NV but far from the target nucleus. Remarkably, we find the $^1$H spin can still polarize efficiently to about 50% of the ideal value, provided the effective bystander spin recycling time $T_1^{(B1)}$ (defined by spin-lattice relaxation or spin diffusion) satisfies $\tau_{LZ} < T_1^{(B1)} \lesssim T_c/2 \ll T_1^{(1H)}$, with $T_1^{(1H)}$ denoting the $^1$H nuclear spin-lattice relaxation time. Since the starting distance between the NV and proxy spin is 5 nm, the presence of an additional bystander P1 at ~3.6 nm from the NV is only likely at P1 concentrations of ~50 ppm or greater, meaning that the technique is expected to work reasonably well in representative diamond surfaces. We refer the reader to the Supplementary Information, Section VII, for further details on these calculations.

In summary, cross-relaxation of the NV center spin with surrounding paramagnetic impurities at low magnetic fields can be exploited to induce spin polarization of nuclear targets not interacting with the NV. Central to this approach is the Landau-Zener dynamics induced by partly non-adiabatic magnetic field sweeps across the set of avoided crossings from nearly-matched energy differences of the individual NV and P1 spins. Somewhat counterintuitively, our work shows that successive field sweeps in opposite directions contribute constructively to the DNP process, to ultimately yield a net nuclear polarization comparable to that of the NV spin, with a sign defined by the relative timing of the optical excitation. This proxy-mediated DNP mechanism does not require the use of microwave, should operate under ambient conditions, and is robust to spin coupling heterogeneity and NV orientation disorder. Unlike prior demonstrations of P1-assisted NV-DNP, these traits make this approach applicable to diamond powders, and insensitive to magnetic field heterogeneity (both spatial and temporal), or system fluctuations (induced, e.g., by laser heating). Interestingly, our calculations indicate that spin-lattice relaxation of the proxy spin can have a positive impact on the DNP efficiency, either by broadening the range of sweep velocities where the transfer remains efficient, and/or by mitigating the adverse effect of strong decoherence (i.e., the regime where $T_2^{(P1)} \ll \tau_{LZ}$). A table with a summary on the range of conditions where this technique is expected to work well can be found in the Supplementary Information, Section VIII.

Because the present technique promises to remain effective even for weak spin couplings, we anticipate proxy-mediated DNP can transfer polarization directly to nuclear spin targets outside the diamond crystal. We contrast this mechanism to nuclear-spin-diffusion transfer, inherently slower and thus more sensitive to shallow-defect-induced spin-lattice relaxation. Finally, we anticipate several extensions of the present technique, for example, in the form of double-resonance schemes at low magnetic fields (e.g., ~10 mT) designed to recreate analogous three-spin-LZ-dynamics in the rotating frame. Potential advantages include the ability to access all NV orientations without compromising on the NV spin pumping efficiency, and the option to separately optimize the sweep velocity and repetition rates via the use of frequency combs[31].

## ASSOCIATED CONTENT

**Supporting Information:** Contains information on the Hamiltonian model, the dynamics of single sweeps, the effect of transversal and longitudinal relaxation, a description of the multi-cycle dynamics via a transfer matrix formalism, the impact of additional NV-coupled P1 centers on the polarization transfer efficiency, and a summary of the protocol's main features and operating conditions.

## AUTHOR INFORMATION

**Corresponding author:**
[†]E-mail: cmeriles@ccny.cuny.edu
**Notes**
The authors declare no competing financial interests.

## ACKNOWLEDGMENTS

P.R.Z., J.H., D.P. and C.A.M. acknowledge support from the National Science Foundation through grants NSF-1619896 and NSF-1401632, and from Research Corporation for Science Advancement through a FRED Award; they also acknowledge access to the facilities and research infrastructure of the NSF CREST IDEALS, grant number NSF-HRD-1547830. J.H. acknowledges support from CREST-PRF NSF-HRD 1827037. All authors acknowledge the CUNY High Performance Computing Center (HPCC). The CUNY HPCC is operated by the College of Staten Island and funded, in part, by grants from the City of New York, State of New York, CUNY Research Foundation, and National Science Foundation Grants CNS-0958379, CNS-0855217 and ACI 1126113.

# Supporting information

## Two-electron-spin ratchets as a platform for microwave-free dynamic nuclear polarization of arbitrary material targets

Pablo R. Zangara[1], Jacob Henshaw[1], Daniela Pagliero[1], Ashok Ajoy[3], Jeffrey A. Reimer[4], Alexander Pines[3], and Carlos A. Meriles[1,2,†]

[1]Dept. of Physics, CUNY-City College of New York, New York, NY 10031, USA.

[2]CUNY-Graduate Center, New York, NY 10016, USA.

[3]Department of Chemistry, University of California Berkeley, and Materials Science Division Lawrence Berkeley National Laboratory, Berkeley, California 94720, USA.

[4]Department of Chemical and Biomolecular Engineering, and Materials Science Division Lawrence Berkeley National Laboratory University of California, Berkeley, California 94720, USA.

### S.I The Hamiltonian model

Let us consider the following Hamiltonian:

$$H = H_{NV} + H_{P1} + H_H + H_N + H_{N'} + H_{dip}(\mathbf{S},\mathbf{S}') + H_{dip}(\mathbf{I},\mathbf{S}') + \mathbf{S} \cdot \mathbf{A}_N \cdot \mathbf{K} + \mathbf{S}' \cdot \mathbf{A}_{N'} \cdot \mathbf{K}' \quad (S.1)$$

Here, the spin operators $\mathbf{S}, \mathbf{S}', \mathbf{I}, \mathbf{K}, \mathbf{K}'$ correspond to the NV electronic spin, the P1 electronic spin, the proton nuclear spin, the host $^{14}$N nuclear spin of the NV, and the host $^{14}$N nuclear spin of the P1, respectively. The term $H_{NV}$ stands for the NV Hamiltonian expressed as the sum of the crystal field $D(S^z)^2$ and the Zeeman interaction $-\gamma_e \mathbf{B} \cdot \mathbf{S}$. Notice that our reference frame has the $z$-direction parallel to the crystalline field of the NV center, whose zero-field splitting is given by D. The terms $H_{P1}$ and $H_H$ stand for the corresponding P1 and proton Zeeman interactions, $-\gamma_e \mathbf{B} \cdot \mathbf{S}'$ and $-\gamma_H \mathbf{B} \cdot \mathbf{I}$, respectively. Analogously, $H_N = -\gamma_N \mathbf{B} \cdot \mathbf{K} + \mathbf{K} \cdot \mathbf{Q} \cdot \mathbf{K}$ corresponds to the Zeeman and quadrupolar terms for the host $^{14}$N nuclear spin of the NV; a similar expression holds for $H_{N'}$. In addition, $\mathbf{A}_N$ ($\mathbf{A}_{N'}$) is the hyperfine coupling tensor between the NV (P1) electronic spin and the $^{14}$N nuclear spin $\mathbf{K}$ ($\mathbf{K}'$) of the nitrogen host. The standard dipolar interaction is given by

$$H_{dip}(\mathbf{S},\mathbf{S}') = J_{dip}^{S-S'}\left[g_0(\vartheta,\varphi)\left(S^z S'^z - \tfrac{1}{4}S^- S'^+ - \tfrac{1}{4}S^+ S'^-\right) + g_1(\vartheta,\varphi)(S^+ S'^z + S^z S'^+) + \right.$$
$$\left. g_1(\vartheta,-\varphi)(S^- S'^z + S^z S'^-) + g_2(\vartheta,\varphi)S^+ S'^+ + g_2(\vartheta,-\varphi)S^- S'^-\right] \quad (S.2)$$

with

$$J_{dip}^{S-S'} = \frac{\gamma_S \gamma_{S'} \hbar^2}{r_{S,S'}^3} \quad (S.3)$$

$$g_0(\vartheta,\varphi) = (1 - 3(\cos\vartheta)^2) \quad (S.4)$$

$$g_1(\vartheta,\varphi) = -\tfrac{3}{2}\sin(\vartheta)\cos(\vartheta)\, e^{-i\varphi} \quad (S.5)$$

$$g_2(\vartheta, \varphi) = -\frac{3}{4}(\sin(\vartheta))^2 e^{-2i\varphi} \qquad (S.6)$$

Here, $r_{S,S'}$ stands for the distance between the spins $S$ and $S'$, and $(\vartheta, \varphi)$ define the angular orientation of the inter-spin axis. A similar expression applies to the dipolar interaction between the P1 electron spin and the $^1$H spin; we assume the proton is closer to the P1, and correspondingly neglect the dipolar coupling with the NV. In what follows we use $(\vartheta_1, \varphi_1)$ to denote the orientation of the NV-P1 interparticle axis and $(\vartheta_2, \varphi_2)$ to denote the orientation of the H-P1 interparticle axis. Additionally, the magnetic field $\mathbf{B}$ is parameterized as $\mathbf{B} = B(\sin\theta\cos\phi, \sin\theta\sin\phi, \cos\theta)$.

## S.II Gap estimates

We can calculate estimates for the relevant energy gaps in the simplified case where we disregard the terms $H_{N(N')}$ and those representing the hyperfine couplings between the electronic spins and the nuclear hosts. Moreover, we focus on the aligned case $\theta = 0$, $\phi = 0$, so here we consider the Hamiltonian:

$$H^{[\theta=0,\phi=0]} = D(S^z)^2 - \gamma_e B S^z - \gamma_e B S'^z - \gamma_H B I^z + H_{dip}(\mathbf{S},\mathbf{S}') + H_{dip}(\mathbf{I},\mathbf{S}') \qquad (S.7)$$

We start by evaluating $H^{[\theta=0,\phi=0]}$ in the Hilbert subspace spanned by the basis states $\{|0,+\frac{1}{2},\uparrow\rangle, |0,+\frac{1}{2},\downarrow\rangle, |-1,-\frac{1}{2},\uparrow\rangle, |-1,-\frac{1}{2},\downarrow\rangle,\}$ where we label spin states according to the notation $|NV, P1, H\rangle$; in addition, we define $\omega_{0S} = |\gamma_e|B$ and $\omega_{0I} = \gamma_H B$. Then, the matrix representation of $H^{[\theta=0,\phi=0]}$ in this subspace is:

$$H^{[\theta=0,\phi=0]} = \begin{array}{c|cccc} & |0,+\frac{1}{2},\uparrow\rangle & |0,+\frac{1}{2},\downarrow\rangle & |-1,-\frac{1}{2},\uparrow\rangle & |-1,-\frac{1}{2},\downarrow\rangle \\ \hline \langle 0,+\frac{1}{2},\uparrow| & \frac{\omega_{0S}}{2} - \frac{\omega_{0I}}{2} + \frac{Z_2}{4} & V_{SS} & V_{DQ} & 0 \\ \langle 0,+\frac{1}{2},\downarrow| & V_{SS}^\dagger & \frac{\omega_{0S}}{2} + \frac{\omega_{0I}}{2} - \frac{Z_2}{4} & 0 & V_{DQ} \\ \langle -1,-\frac{1}{2},\uparrow| & V_{DQ}^\dagger & 0 & -\frac{3\omega_{0S}}{2} - \frac{\omega_{0I}}{2} + D - \frac{Z_2}{4} - \frac{Z_1}{2} & V_{SS} \\ \langle -1,-\frac{1}{2},\downarrow| & 0 & V_{DQ}^\dagger & V_{SS}^\dagger & -\frac{3\omega_{0S}}{2} + \frac{\omega_{0I}}{2} + D + \frac{Z_2}{4} + \frac{Z_1}{2} \end{array} \qquad (S.8)$$

where

$$Z_1 = g_0(\vartheta_1, \varphi_1) J_{dip}^{NV-P1} \qquad (S.9)$$

$$Z_2 = g_0(\vartheta_2, \varphi_2) J_{dip}^{H-P1} \qquad (S.10)$$

$$V_{SS} = g_1(\vartheta_2, \varphi_2) J_{dip}^{H-P1} \qquad (S.11)$$

$$V_{DQ} = g_2(\vartheta_1, \varphi_1) J_{dip}^{NV-P1} \qquad (S.12)$$

By tuning $B$ one can force the states $|0,+\frac{1}{2},\downarrow\rangle$ and $|-1,-\frac{1}{2},\uparrow\rangle$ to be almost degenerate

$$\frac{\omega_{0S}}{2} + \frac{\omega_{0I}}{2} - \frac{Z_2}{4} \approx -\frac{3\omega_{0S}}{2} - \frac{\omega_{0I}}{2} + D - \frac{Z_2}{4} - \frac{Z_1}{2}, \qquad (S.13)$$

or, equivalently,

$$2\omega_{0S} + \omega_{0I} \approx D - \frac{Z_1}{2}, \tag{S.14}$$

In this condition, the energy of the states $|0, +\frac{1}{2}, \downarrow\rangle$ and $|-1, -\frac{1}{2}, \uparrow\rangle$ is:

$$E_0 = \langle 0, +\tfrac{1}{2}, \downarrow | H^{[\theta=0,\phi=0]} | 0, +\tfrac{1}{2}, \downarrow \rangle = \langle -1, -\tfrac{1}{2}, \uparrow | H^{[\theta=0,\phi=0]} | -1, -\tfrac{1}{2}, \uparrow \rangle \tag{S.15}$$

$$E_0 = \frac{D}{4} + \frac{\omega_{0I}}{4} - \frac{Z_1}{8} - \frac{Z_2}{4} \tag{S.16}$$

In addition,

$$E_a = \langle 0, +\tfrac{1}{2}, \uparrow | H^{[\theta=0,\phi=0]} | 0, +\tfrac{1}{2}, \uparrow \rangle = \frac{D}{4} - \frac{3\omega_{0I}}{4} - \frac{Z_1}{8} + \frac{Z_2}{4} \tag{S.17}$$

$$E_b = \langle -1, -\tfrac{1}{2}, \downarrow | H^{[\theta=0,\phi=0]} | -1, -\tfrac{1}{2}, \downarrow \rangle = \frac{D}{4} + \frac{5\omega_{0I}}{4} + \frac{7Z_1}{8} + \frac{Z_2}{4} \tag{S.18}$$

This yields

$$H^{[\theta=0,\phi=0]} = \begin{array}{c|cccc} & |0,+\tfrac{1}{2},\uparrow\rangle & |0,+\tfrac{1}{2},\downarrow\rangle & |-1,-\tfrac{1}{2},\uparrow\rangle & |-1,-\tfrac{1}{2},\downarrow\rangle \\ \hline \langle 0,+\tfrac{1}{2},\uparrow| & E_a & V_{SS} & V_{DQ} & 0 \\ \langle 0,+\tfrac{1}{2},\downarrow| & V_{SS}^\dagger & E_0 & 0 & V_{DQ} \\ \langle -1,-\tfrac{1}{2},\uparrow| & V_{DQ}^\dagger & 0 & E_0 & V_{SS} \\ \langle -1,-\tfrac{1}{2},\downarrow| & 0 & V_{DQ}^\dagger & V_{SS}^\dagger & E_b \end{array} \tag{S.19}$$

Rather than computing a value of the energy gap that opens at the degeneracy point, we need to identify an effective coupling between the states $|0, +\frac{1}{2}, \downarrow\rangle$ and $|-1, -\frac{1}{2}, \uparrow\rangle$. Such a coupling matrix element is the one responsible for the LZ dynamics. Thus, we use a second order formula to estimate a virtual interaction element:

$$J_v^{(a)} = \frac{\langle -1,-\tfrac{1}{2},\uparrow | H^{[\theta=0,\phi=0]} | 0,+\tfrac{1}{2},\uparrow \rangle \langle 0,+\tfrac{1}{2},\uparrow | H^{[\theta=0,\phi=0]} | 0,+\tfrac{1}{2},\downarrow \rangle}{\langle 0,+\tfrac{1}{2},\downarrow | H^{[\theta=0,\phi=0]} | 0,+\tfrac{1}{2},\downarrow \rangle - \langle -1,-\tfrac{1}{2},\downarrow | H^{[\theta=0,\phi=0]} | -1,-\tfrac{1}{2},\downarrow \rangle} \tag{S.20}$$

$$J_v^{(a)} = \frac{V_{DQ}^\dagger V_{SS}}{E_0 - E_a} = \frac{V_{DQ}^\dagger V_{SS}}{\omega_{0I} - \frac{Z_2}{2}} \tag{S.21}$$

An alternative option for this virtual process is given by a different intermediate state:

$$J_v^{(b)} = \frac{\langle -1,-\tfrac{1}{2},\uparrow | H^{[\theta=0,\phi=0]} | -1,-\tfrac{1}{2},\downarrow \rangle \langle -1,-\tfrac{1}{2},\downarrow | H^{[\theta=0,\phi=0]} | 0,+\tfrac{1}{2},\downarrow \rangle}{\langle 0,+\tfrac{1}{2},\downarrow | H^{[\theta=0,\phi=0]} | 0,+\tfrac{1}{2},\downarrow \rangle - \langle -1,-\tfrac{1}{2},\downarrow | H^{[\theta=0,\phi=0]} | -1,-\tfrac{1}{2},\downarrow \rangle} \tag{S.22}$$

$$J_v^{(b)} = \frac{V_{DQ}^\dagger V_{SS}}{E_0 - E_b} = -\frac{V_{DQ}^\dagger V_{SS}}{\omega_{0I} + Z_1 + \frac{Z_2}{2}} \tag{S.23}$$

Adding the two contributions,

$$J_{virtual} = J_v^{(a)} + J_v^{(b)} = V_{DQ}^\dagger V_{SS} \frac{(Z_1+Z_2)}{\left(\omega_{0I}-\frac{Z_2}{2}\right)\left(\omega_{0I}+Z_1+\frac{Z_2}{2}\right)} \quad (S.24)$$

This virtual or second order interaction yields the estimated transition element $\Delta_1 \sim 2|J_{virtual}|$ that corresponds to the gap at the degeneracy point between the states $|0,+\frac{1}{2},\downarrow\rangle$ and $|-1,-\frac{1}{2},\uparrow\rangle$. The same transition element is expected at the degeneracy point between the states $|0,+\frac{1}{2},\uparrow\rangle$ and $|-1,-\frac{1}{2},\downarrow\rangle$. In addition, $\Delta_1$ has to be compared with the much bigger -direct- gap $\Delta_0 \sim 2|V_{DQ}|$ that opens between the states $|0,+\frac{1}{2},\downarrow\rangle$ and $|-1,-\frac{1}{2},\downarrow\rangle$ and between the states $|0,+\frac{1}{2},\uparrow\rangle$ and $|-1,-\frac{1}{2},\uparrow\rangle$ (see Fig. 1(d) – main text).

## S.III Single passage and nitrogen hosts

Performing a field sweep in a given direction drives the spin populations into a series of LZ transitions. As already mentioned, the two $\Delta_1$-gaps are crossed as we sweep through the two degeneracy points (between $|0,+\frac{1}{2},\downarrow\rangle$ and $|-1,-\frac{1}{2},\uparrow\rangle$ and between $|0,+\frac{1}{2},\uparrow\rangle$ and $|-1,-\frac{1}{2},\downarrow\rangle$). The presence of two successive LZ bifurcations produces the well-known Stuckelberg interferences, which modulate the obtained polarization as a function of the sweep velocity[1] $\beta$. This strong dependence on $\beta$ is explicitly shown in Fig. S1-(**a**). The optimal rate $\beta$ that yields the maximal polarization in a single sweep ultimately depends on the precise value of the coupling strengths in Eq. (S.7). Figure S1-(**b**) shows that similar levels of proton polarization can be achieved when the nitrogen hosts are included, i.e. the terms $H_{N(N')}$ and the corresponding hyperfine couplings in Eq. (S.1). In this case, though, reaching the maximal polarization is subtler since we need to sweep different regions at different $\beta$s. Thus, the presence of the nitrogen hosts does not introduce any fundamental change to the dynamics we are trying to model. Since it increases both the complexity of the energy diagrams and the simulation time, we ignore the nuclear hosts in our simulations.

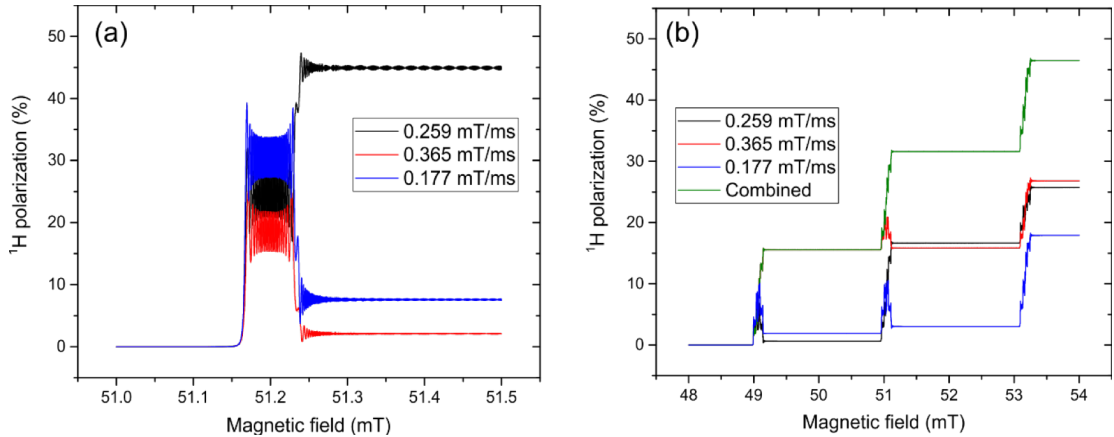

**Figure S1**. Comparison of single low-to-high sweeps at different velocities $\beta$, with and without nitrogen hosts. In all cases a low-to-high field sweep is performed with the magnetic field aligned to the NV axis, i.e. $\theta = 0, \phi = 0$. The NV at $t = 0$ is fully polarized. $J_{dip}^{NV-P1} = 500$ kHz, $J_{dip}^{H-P1} = 200$ kHz. (**a**) No nuclear hosts considered. (b) Nitrogen hosts included; $\|A_N\| \sim 2$ MHz, $\|A_{N'}\| \sim 115$ MHz. The 'combined' case (green line) involves sweeping the first manifold ($m_K' = -1$, range 48 mT $\to$ 50 mT) at $\beta = 0.365$ mT/ms, the second manifold ($m_K' = 0$, range 50 mT $\to$ 52 mT) at $\beta = 0.259$ mT/ms, and the third manifold ($m_K' = +1$, range 52 mT $\to$ 54 mT) at $\beta = 0.177$ mT/ms.

## S.IV Ratchet dynamics

We represent the light pulse by a deterministic projection of the NV state (reduced density matrix) into

$$\rho_i^{NV} = \frac{(1+\epsilon)}{3}|0\rangle\langle 0| + \frac{(1-\epsilon/2)}{3}(|-1\rangle\langle -1| + |+1\rangle\langle +1|) \quad (S.25)$$

Here, the parameter $\epsilon$ controls the degree of the light-induced spin initialization. More precisely, $\epsilon = \epsilon_0 \eta_{NV}$, where $\epsilon_0$ is, in general, a function of the intensity and duration of the light pulse, and $\eta_{NV}$ the NV initialization efficiency, which depends on the magnetic field misalignment (see Fig. 4(c) – main text). In our calculations, we model optical NV repolarization by expressing the total density matrix $\rho$ in the form $\rho_i^{NV} \otimes Tr_{NV}\{\rho\}$, where $Tr_{NV}\{\cdot\}$ denotes the partial trace over the NV degrees of freedom.

In all simulations reported herein we assume that the cycle time $T_C$ is always shorter than (or at least comparable to) the NV $T_1$ time. For instance, the longest $T_C$ considered in Fig. 3 (main text) is 1 ms (this case corresponds to $\beta_{up} = \beta_{down} = 1$ mT/ms and a field range of 0.5 mT). This ensures that the LZ crossings always occur with the NV polarized. In the extreme case where the cycle time $T_C$ is much shorter than the NV $T_1$ time, the repolarization pulse does not need to be present at every cycle. In particular, we explore in Fig. S2-(**a-b**) this possibility. By assuming the NV $T_1$ time to be sufficiently long, we repolarize the spin state of the NV into $m_S = 0$ only after $l = 1, 2, 5, 10$ cycles. In some cases, the phase coherence between consecutive cycles can lead to a strongly non-monotonic behavior (Fig. S2(**a**)).

Erasing the phase coherence of the system corresponds to introducing $T_2$ relaxation. In practice, we force such a dephasing by making zero the off-diagonal elements of the complete density matrix written in the instantaneous eigenbasis. Thus, only the populations remain unchanged. In principle, the dephasing does not alter substantially the efficiency in the generation of polarization, but specific interferences are lost. For instance, if we include dephasing after sweeping across the resonance, the dynamics of the cycles yield a monotonic, cumulative increase, even in the case $l = 10$, see Fig. S2-(**b**). Additionally, we show in Fig. S2-(**c**) that the absence of phase coherence between two consecutive passages produces an approximately uniform height of each step in the buildup of polarization (when coherence is maintained, the two step-like jumps per cycle need not be equal; in the case shown, the sweep down produces a higher increase).

In order to systematically account for decoherence in our simulations, we impose dephasing at the end of each sweep (both up and down, i.e. two dephasing operations per cycle) and keep the LZ dynamics coherent. The precise time when we perform this operation does not alter the generated polarization, provided that it does not happen during the LZ crossings. In the limit of strong dephasing and low velocities (e.g., $\beta_{up}, \beta_{down} \lesssim 5$ mT/ms), decoherence will certainly occur during the LZ crossings (particularly across the wide $\Delta_0$-gap). We address this specific case in Sec. S.VI using the framework introduced in Sec. S.V.

The effect of the longitudinal relaxation of the P1 spin is indeed quite important (compare Figs. 3-(**b**) and 3-(**c**), main text). In general, we observe that the mechanism of polarization transfer is more efficient when $T_1$ relaxation is included as compared to the case where the $T_1$ time is assumed infinite. This comparison is illustrated in Fig. S3. In particular, we show in Fig. S3-(**a**) the evolution of the polarization, and in Figs. S3-(**b**) and S3-(**c**) the evolution of populations of the four $m_S = 0$ states. It is clear that, in the case where $T_1$ relaxation is included (Fig. S3-(**c**)), every cycle steadily increases the population imbalance between the proton spin projections ↑ and ↓, regardless of the spin state of the P1. This is not the case when $T_1$ relaxation is not included (Fig. S3-(**b**)).

A simple pictorial approach can be used to understand the basic role of spin-lattice relaxation in enhancing the generation of proton polarization. Figures S4 and S5 show a series of energy level diagrams with a schematic representation of population dynamics. In Fig. S4, which represents the dynamics without spin-lattice relaxation, we show that after some polarization is created (first cycle) all populations are pumped into the subspace $m_s = 0$, $m'_S = -1/2$ where they remain trapped. Thus, any successive cycle cannot create more polarization. Quite on the contrary, Fig. S5 shows that spin-lattice relaxation for the P1 center can spread the population imbalance across a larger (unconstrained) state space, leading to a more robust and additive generation of nuclear polarization. We stress that in this euristic approach no effect of decoherence is considered. A more involved analysis based on a Transfer Matrix (TM) technique leads to the same conclusions (Sections S.V and S.VI).

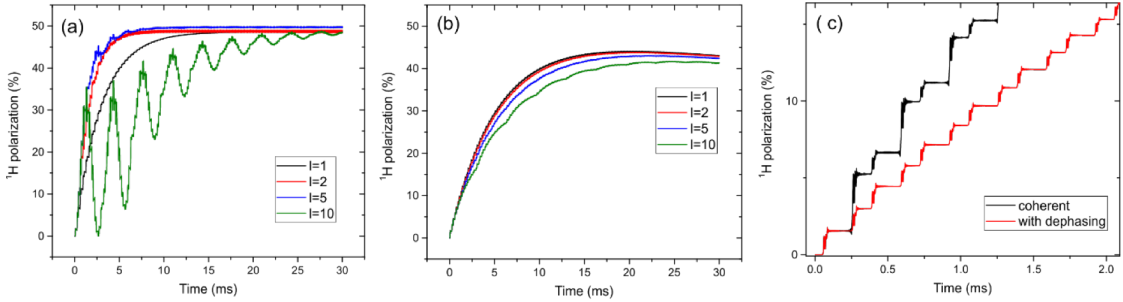

**Figure S2**. Multiple cycles and phase coherence. In all cases the magnetic field is aligned with the NV axis, i.e. $\theta = 0$, $\phi = 0$. Every time the NV is initialized, its spin is polarized up to 100% ($\epsilon = 2.0$). Initialization always takes place at the beginning of the cycle (as in Fig. 2(**a**), main text). $J_{dip}^{NV-P1} = 500$ kHz, $J_{dip}^{H-P1} = 100$ kHz, $\beta_{up} = \beta_{down} = 3$ mT/ms. (**a**) Coherent case, no $T_2$ dephasing is considered. (**b**) Two dephasing events per cycle, one after every sweep (up and down). In both (**a**) and (**b**) the insets indicate the number of cycles $l$ per NV initialization. (**c**) First few cycles for both the coherent and with-dephasing cases, with NV initialization for every cycle ($l = 1$).

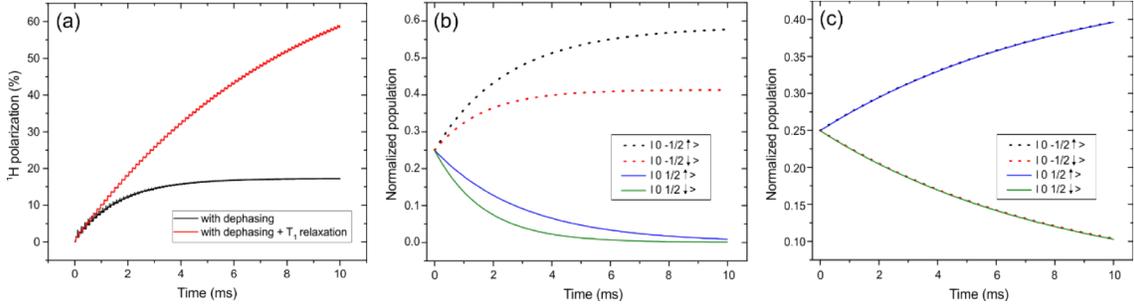

**Figure S3.** The effect of longitudinal relaxation for the P1 spin. We assume the magnetic field to be aligned with the NV axis, i.e. $\theta = 0$, $\phi = 0$. In order to exemplify the dynamics of populations, every time the NV is initialized, its spin is polarized up to 100% ($\epsilon = 2.0$). The initialization always happens at the beginning of the cycle (as in Fig. 2(**a**), main text). $J_{dip}^{NV-P1} = 500$ kHz, $J_{dip}^{H-P1} = 100$ kHz, $\beta_{up} = 6$ mT/ms and $\beta_{down} = 10$ mT/ms. (**a**) The evolution of proton polarization. In (**b-c**) we show the corresponding dynamics of populations in the four $m_s = 0$ states at the beginning of each cycle (all populations have been projected into this subspace by the NV initialization). (**b**) With dephasing. (**c**) With dephasing and $T_1$ relaxation for the P1.

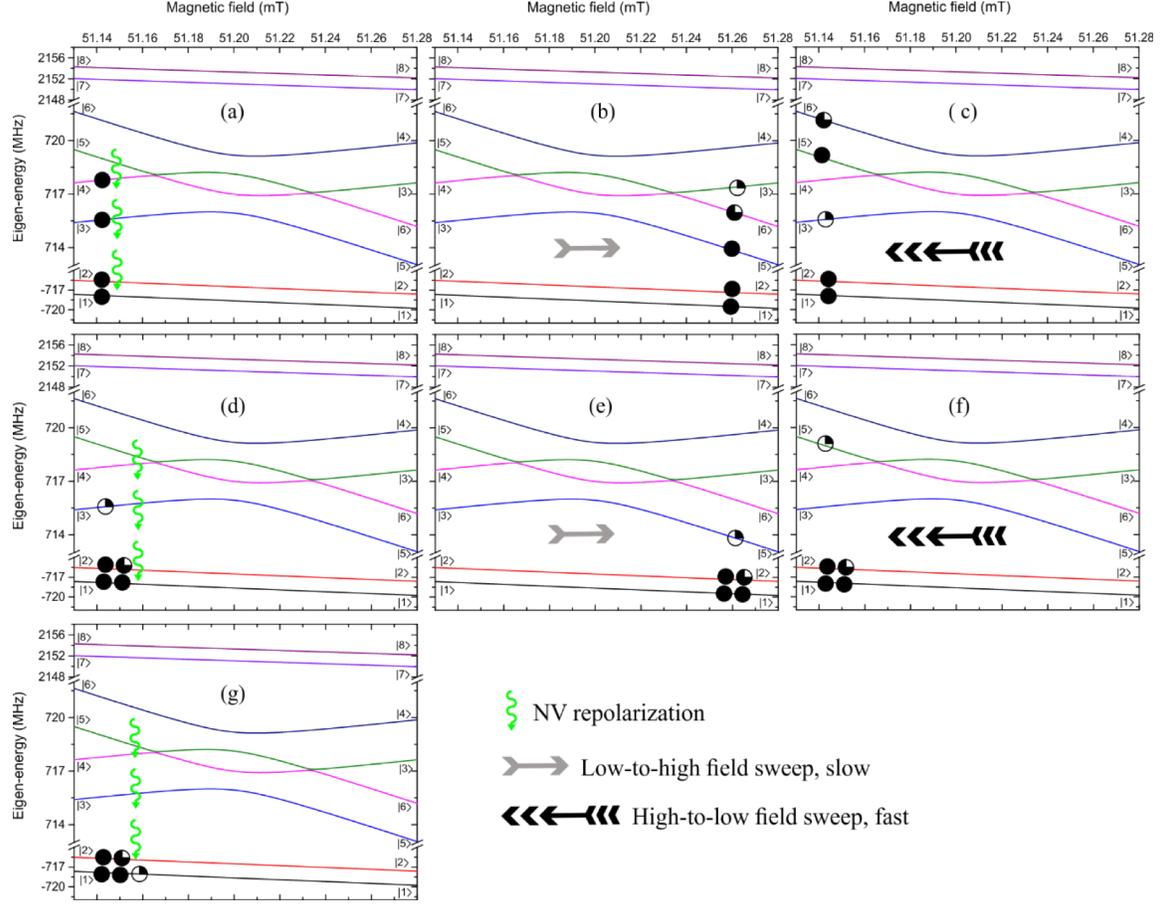

**Figure S4**. Population dynamics without spin-lattice relaxation. Each diagram shows the eigen-energies as function of the magnetic field, with the labelling as follows: $|1\rangle = |0, -\frac{1}{2}, \uparrow\rangle$, $|2\rangle = |0, -\frac{1}{2}, \downarrow\rangle$, $|3\rangle = |0, \frac{1}{2}, \uparrow\rangle$, $|4\rangle = |0, \frac{1}{2}, \downarrow\rangle$, $|5\rangle = |-1, -\frac{1}{2}, \uparrow\rangle$, $|6\rangle = |-1, -\frac{1}{2}, \downarrow\rangle$, $|7\rangle = |-1, \frac{1}{2}, \uparrow\rangle$, $|8\rangle = |-1, \frac{1}{2}, \downarrow\rangle$. Spin populations are represented by spheres and the variable filling indicates population imbalance. (**a**) Initially the states with $m_s = 0$ ($|1\rangle, |2\rangle, |3\rangle$, and $|4\rangle$) are equally populated after the light initialization. (**b**) A moderately slow low-to-high sweep generates polarization. (**c**) A very fast high-to-low sweep drives the populations non-adiabatically, keeping them in the same state. (**d**) Light repolarization occurs, bringing most of the population into the subspace with $m_s = 0$, $m_S' = -1/2$. This completes the first cycle. The following cycle (**e-g**) ends up dragging the remaining population into the same subspace, without creating any further population imbalance (proton polarization).

### S.V Transfer Matrix model

In order to describe the DNP cycle dynamics using a TM approach, we first consider the probability $p_0$ of crossing non-adiabatically a $\Delta_0$-gap. This probability equals to ~1 only in the limit of complete non-adiabaticity (very fast sweep velocity) and it can be written in terms of the standard LZ formula[1]:

$$p_0^{\nearrow} \sim exp\left\{-\frac{\pi}{2} \times \frac{\Delta_0^2}{2|\gamma_e|\beta_{up}}\right\} \qquad (S.26)$$

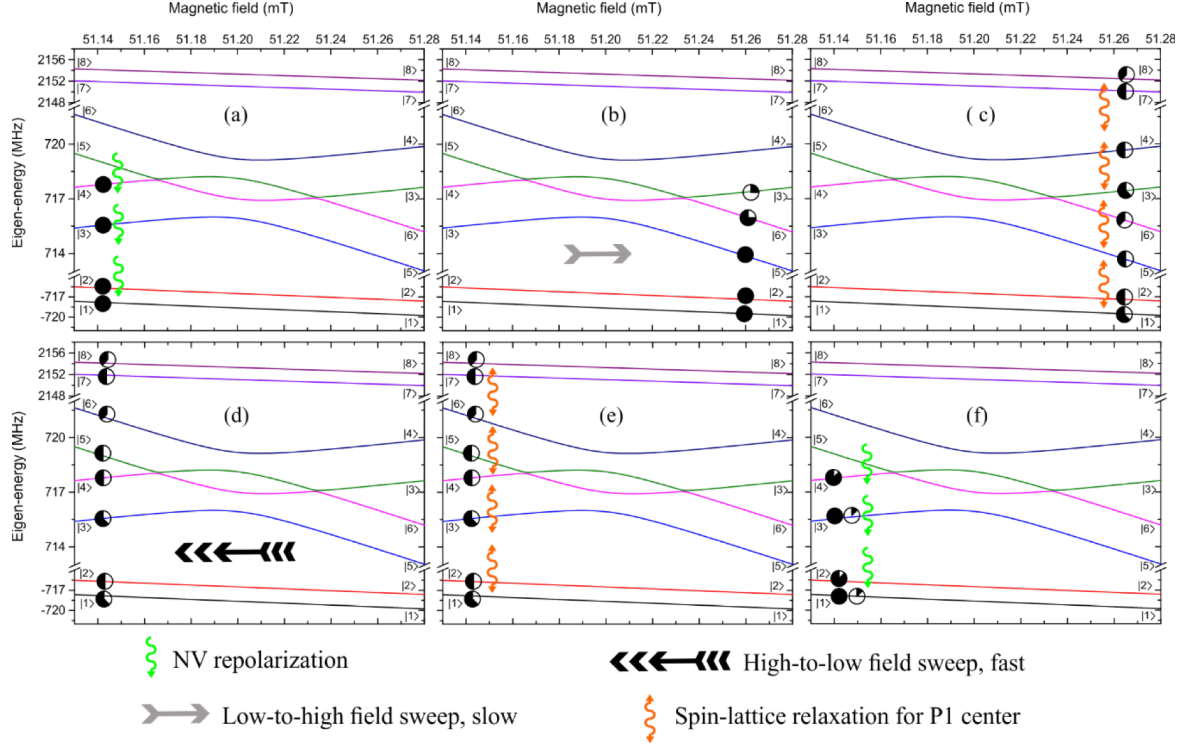

**Figure S5**. Population dynamics with spin-lattice relaxation, same notation as in Fig. S4. (**a**) Initially the states with $m_s = 0$ ($|1\rangle$, $|2\rangle$, $|3\rangle$, and $|4\rangle$) are equally populated after the light initialization. (**b**) A moderately slow low-to-high sweep generates polarization. (**c**) Spin-lattice relaxation for the P1 center forces equal population between the states: $|1\rangle$ and $|3\rangle$; $|2\rangle$ and $|4\rangle$; $|5\rangle$ and $|7\rangle$; $|6\rangle$ and $|8\rangle$. (**d**) A very fast high-to-low sweep drives the populations non-adiabatically, keeping them in the same state. (**e**) Spin-lattice relaxation occurs. This step may not happen at all, but even if it happens, it does not introduce any change. (**f**) Light repolarization occurs, bringing all the population into the subspace with $m_s = 0$. Notice that, in the end, the same population imbalance exists between states $|1\rangle$ and $|2\rangle$ and between $|3\rangle$ and $|4\rangle$. Any subsequent cycle will symmetrically enlarge this imbalance.

The superscript ↗ indicates the direction of the sweep (low-to-high), the factor $2|\gamma_e|$ in the denominator comes from the time derivative of the energy, and we assume some arbitrary (but fixed) sweep rate $\beta_{up}$. Analogously, the LZ formula for the probability of a non-adiabatic crossing at a $\Delta_1$-gap is given by:

$$p_1^\nearrow \sim exp\left\{-\frac{\pi}{2} \times \frac{\Delta_1^2}{(2|\gamma_e|+\gamma_H)\beta_{up}}\right\} \quad (S.27)$$

We can now use these probabilities to determine the different TM elements: Following the state notation introduced in Fig. S4, if the system is initially in state $|3\rangle$, it can undergo a fully adiabatic evolution with probability $1 - p_0^\nearrow$ and end up in state $|5\rangle$. Alternatively, there is a probability $p_0^\nearrow p_1^\nearrow$ that after the sweep it remains in state $|3\rangle$, which corresponds to a full non-adiabatic crossing of both gaps ($\Delta_0$ and $\Delta_1$). In addition, there is a nonzero (but very small) probability $p_0^\nearrow(1 - p_1^\nearrow)$ associated with the population ending up in state $|6\rangle$. Analogously, if the system is initially in state $|6\rangle$, then it will end up with probability $1 - p_0^\nearrow$ in $|4\rangle$, with probability $p_0^\nearrow p_1^\nearrow$ in $|6\rangle$, and with probability $p_0^\nearrow(1 - p_1^\nearrow)$ in $|3\rangle$.

The analysis for $|4\rangle$ or $|5\rangle$ as initial states is slightly different. As before, if the system is initially in state $|4\rangle$ then there is a probability $p_1^\nearrow p_0^\nearrow$ that after the sweep it remains in $|4\rangle$ (non-adiabatic limit). In addition, if the state does change, the system may end up in states $|3\rangle$, $|6\rangle$, or even in $|5\rangle$ (unlikely) depending upon the dynamics across the consecutive branchings. Taking into account the conditional probability $(1 - p_0^\nearrow)$ and the two $\Delta_1$-bifurcations, we can estimate the probabilities of ending in state $|3\rangle$ or $|6\rangle$ for a system that was initially in state $|4\rangle$:

$$p_{|4\rangle \to |3\rangle}^\nearrow = (1 - p_0^\nearrow) \times 2p_1^\nearrow(1 - p_1^\nearrow), \tag{S.28}$$

$$p_{|4\rangle \to |6\rangle}^\nearrow = (1 - p_0^\nearrow) \times [p_1^{\nearrow 2} + (1 - p_1^\nearrow)^2] \tag{S.29}$$

The small probability $p_{|4\rangle \to |5\rangle}^\nearrow$ is given by $p_0^\nearrow(1 - p_1^\nearrow)$. It is clear also that similar arguments can be made for $p_{|5\rangle \to |5\rangle}^\nearrow$, $p_{|5\rangle \to |3\rangle}^\nearrow$, $p_{|5\rangle \to |6\rangle}^\nearrow$ and $p_{|5\rangle \to |4\rangle}^\nearrow$.

It is worth noting here that, even though we are considering the LZ formula (which is a purely quantum-mechanical result), the computation of transfer probabilities is entirely classical, since no interference effects are considered. The appropriate quantum prediction of a multiple passage yields the Stuckelberg interference described in Section S.III.

We can now write down a TM representation for a low-to-high sweep,

|   | $\|1\rangle$ | $\|2\rangle$ | $\|3\rangle$ | $\|4\rangle$ | $\|5\rangle$ | $\|6\rangle$ | $\|7\rangle$ | $\|8\rangle$ |
|---|---|---|---|---|---|---|---|---|
| $\langle 1\|$ | 1 | 0 | 0 | 0 | 0 | 0 | 0 | 0 |
| $\langle 2\|$ | 0 | 1 | 0 | 0 | 0 | 0 | 0 | 0 |
| $\langle 3\|$ | 0 | 0 | $p_0^\nearrow p_1^\nearrow$ | $2p_{0\perp}^\nearrow p_1^\nearrow p_{1\perp}^\nearrow$ | $p_{0\perp}^\nearrow(p_1^{\nearrow 2} + p_{1\perp}^{\nearrow 2})$ | $p_0^\nearrow p_{1\perp}^\nearrow$ | 0 | 0 |
| $\mathbb{T}_\nearrow = \langle 4\|$ | 0 | 0 | 0 | $p_1^\nearrow p_0^\nearrow$ | $p_{1\perp}^\nearrow p_0^\nearrow$ | $p_{0\perp}^\nearrow$ | 0 | 0 |
| $\langle 5\|$ | 0 | 0 | $p_{0\perp}^\nearrow$ | $p_{1\perp}^\nearrow p_0^\nearrow$ | $p_1^\nearrow p_0^\nearrow$ | 0 | 0 | 0 |
| $\langle 6\|$ | 0 | 0 | $p_0^\nearrow p_{1\perp}^\nearrow$ | $p_{0\perp}^\nearrow(p_1^{\nearrow 2} + p_{1\perp}^{\nearrow 2})$ | $2p_{0\perp}^\nearrow p_1^\nearrow p_{1\perp}^\nearrow$ | $p_0^\nearrow p_1^\nearrow$ | 0 | 0 |
| $\langle 7\|$ | 0 | 0 | 0 | 0 | 0 | 0 | 1 | 0 |
| $\langle 8\|$ | 0 | 0 | 0 | 0 | 0 | 0 | 0 | 1 |

where we define $p_{0\perp}^\nearrow = (1 - p_0^\nearrow)$ and $p_{1\perp}^\nearrow = (1 - p_1^\nearrow)$. The corresponding high-to-low TM is obtained by transposition,

|   | $\|1\rangle$ | $\|2\rangle$ | $\|3\rangle$ | $\|4\rangle$ | $\|5\rangle$ | $\|6\rangle$ | $\|7\rangle$ | $\|8\rangle$ |
|---|---|---|---|---|---|---|---|---|
| $\langle 1\|$ | 1 | 0 | 0 | 0 | 0 | 0 | 0 | 0 |
| $\langle 2\|$ | 0 | 1 | 0 | 0 | 0 | 0 | 0 | 0 |
| $\langle 3\|$ | 0 | 0 | $p_1^\searrow p_0^\searrow$ | 0 | $p_{0\perp}^\searrow$ | $p_{1\perp}^\searrow p_0^\searrow$ | 0 | 0 |
| $\mathbb{T}_\searrow = \langle 4\|$ | 0 | 0 | $2p_{0\perp}^\searrow p_1^\searrow p_{1\perp}^\searrow$ | $p_1^\searrow p_0^\searrow$ | $p_{1\perp}^\searrow p_0^\searrow$ | $p_{0\perp}^\searrow(p_1^{\searrow 2} + p_{1\perp}^{\searrow 2})$ | 0 | 0 |
| $\langle 5\|$ | 0 | 0 | $p_{0\perp}^\searrow(p_1^{\searrow 2} + p_{1\perp}^{\searrow 2})$ | $p_{1\perp}^\searrow p_0^\searrow$ | $p_1^\searrow p_0^\searrow$ | $2p_{0\perp}^\searrow p_1^\searrow p_{1\perp}^\searrow$ | 0 | 0 |
| $\langle 6\|$ | 0 | 0 | $p_{1\perp}^\searrow p_0^\searrow$ | $p_{0\perp}^\searrow$ | 0 | $p_1^\searrow p_0^\searrow$ | 0 | 0 |
| $\langle 7\|$ | 0 | 0 | 0 | 0 | 0 | 0 | 1 | 0 |
| $\langle 8\|$ | 0 | 0 | 0 | 0 | 0 | 0 | 0 | 1 |

Similarly, the effect of $T_1$ relaxation in the P1 spin can be represented by another transfer matrix:

$$\mathbb{T}_{T_1} = \begin{array}{c} \\ \langle 1| \\ \langle 2| \\ \langle 3| \\ \langle 4| \\ \langle 5| \\ \langle 6| \\ \langle 7| \\ \langle 8| \end{array} \begin{pmatrix} |1\rangle & |2\rangle & |3\rangle & |4\rangle & |5\rangle & |6\rangle & |7\rangle & |8\rangle \\ 1/2 & 0 & 1/2 & 0 & 0 & 0 & 0 & 0 \\ 0 & 1/2 & 0 & 1/2 & 0 & 0 & 0 & 0 \\ 1/2 & 0 & 1/2 & 0 & 0 & 0 & 0 & 0 \\ 0 & 1/2 & 0 & 1/2 & 0 & 0 & 0 & 0 \\ 0 & 0 & 0 & 0 & 1/2 & 0 & 1/2 & 0 \\ 0 & 0 & 0 & 0 & 0 & 1/2 & 0 & 1/2 \\ 0 & 0 & 0 & 0 & 1/2 & 0 & 1/2 & 0 \\ 0 & 0 & 0 & 0 & 0 & 1/2 & 0 & 1/2 \end{pmatrix}$$

The TM for a light-induced (full-) NV repolarization into the $m_s = 0$ subspace is:

$$\mathbb{T}_L = \begin{array}{c} \\ \langle 1| \\ \langle 2| \\ \langle 3| \\ \langle 4| \\ \langle 5| \\ \langle 6| \\ \langle 7| \\ \langle 8| \end{array} \begin{pmatrix} |1\rangle & |2\rangle & |3\rangle & |4\rangle & |5\rangle & |6\rangle & |7\rangle & |8\rangle \\ 1 & 0 & 0 & 0 & 1 & 0 & 0 & 0 \\ 0 & 1 & 0 & 0 & 0 & 1 & 0 & 0 \\ 0 & 0 & 1 & 0 & 0 & 0 & 1 & 0 \\ 0 & 0 & 0 & 1 & 0 & 0 & 0 & 1 \\ 0 & 0 & 0 & 0 & 0 & 0 & 0 & 0 \\ 0 & 0 & 0 & 0 & 0 & 0 & 0 & 0 \\ 0 & 0 & 0 & 0 & 0 & 0 & 0 & 0 \\ 0 & 0 & 0 & 0 & 0 & 0 & 0 & 0 \end{pmatrix}$$

Therefore, a full cycle including $T_1$ relaxation of the P1 can be modeled as $\mathbb{T}_C^r = \mathbb{T}_L \mathbb{T}_{T_1} \mathbb{T}_\searrow \mathbb{T}_{T_1} \mathbb{T}_\nearrow$.

**S.VI Decoherence and spin-lattice relaxation**

Here, we use the TM approach introduced in Section S.V to analyze specific regimes. First, we consider the particular case when dephasing occurs during the LZ crossings. As mentioned in S.IV, this case in particularly relevant for low velocities (e.g. $\beta_{up}, \beta_{down} \lesssim 5$ mT/ms). In general, decoherence reduces the apparent adiabaticity and, strictly speaking, the complete adiabatic regime is no longer accessible[2-4]. Indeed, in the strong-dephasing (SD) limit, the LZ probability in Eq. (S.26) changes to

$$p_{0,SD}^\nearrow \sim \frac{1}{2}\left[1 + exp\left\{-\pi \times \frac{\Delta_0^2}{2|\gamma_e|\beta_{up}}\right\}\right] \tag{S.30}$$

Therefore, as $\beta_{up} \to 0$ we have $p_{0,SD}^\nearrow \to \frac{1}{2}$, which implies a diffusive scenario where both branches in the LZ crossing end up equally populated. The validity of Eq. (S.30) ultimately depends on the relation between the $T_2$ time and the LZ time $\tau_{LZ} \sim \Delta_0/(2|\gamma_e|\beta_{up})$. In the case of a moderately slow sweep velocity, say $\beta_{up} = 3$ mT/ms, we have $\tau_{LZ} \sim 5$ μs. This means that, for such velocities, even a $T_2$ time of a few microseconds is sufficient to drive the system into the SD limit as it crosses the $\Delta_0$-gap.

In order to address this scenario, we assume $\beta_{up} = \beta_{down}$ and we evaluate the LZ probabilities at the $\Delta_0$-gap in the SD limit,

$$p^\nearrow_{0,SD} = p^\nearrow_{0,\perp,SD} = p^\searrow_{0,SD} = p^\searrow_{0,\perp,SD} = \frac{1}{2}, \quad (S.31)$$

while we keep $p^\nearrow_1 = p^\searrow_1$ very close to one. Figure S6 shows the polarization obtained using the TM approach under these conditions. The proton polarization is systematically larger in the case where spin-lattice relaxation is included as compared to the case where it is not. As a matter of fact, a sweep velocity of $\beta_{up} = \beta_{down} = 3$ mT/ms roughly corresponds to $p^\nearrow_1 = p^\searrow_1 \sim 0.98$, so the presence of spin-lattice relaxation corresponds to an approximate ten-fold enhancement in the proton polarization.

We stress that the SD condition only applies when $T_2/\tau_{LZ} \ll 1$, which for the example given above ($\tau_{LZ} \sim 5$ μs), implies $T_2 \lesssim 1$ μs. Even for $T_2 \sim 100$ ns, a spin-lattice relaxation of $T_1 \sim 100$ $T_2 \sim 10$ μs is sufficient to attain considerable DNP efficiency. Ultimately, we expect a dramatic reduction of the polarization transfer efficiency in the extreme case where $T_1/\tau_{LZ} \ll 1$ (not considered in any of our simulations).

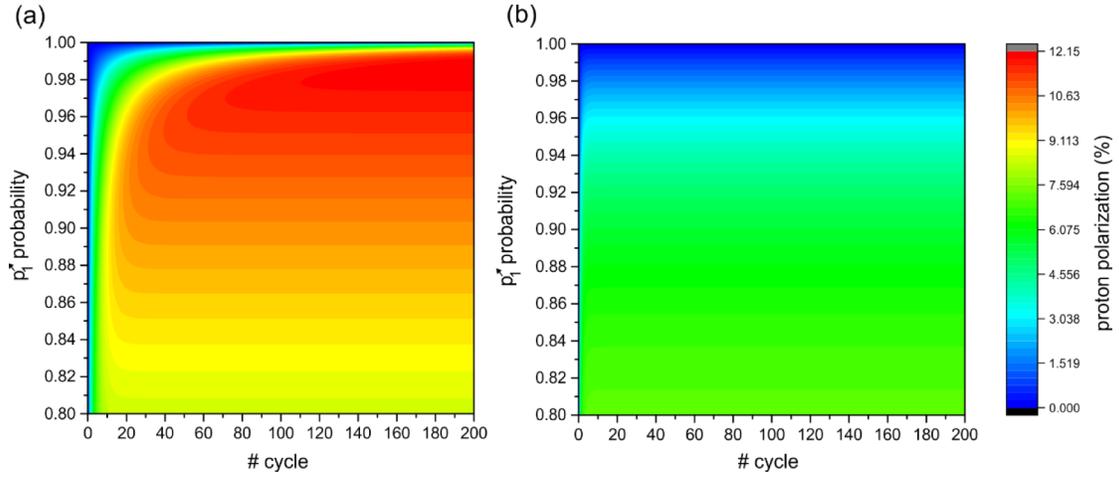

**Figure S6**. Proton polarization evaluated using the TM approach, as a function of the number of cycles and $p^\nearrow_1$. The SD conditions are given by Eq. (S.31). (**a**) With spin-lattice relaxation, $\mathbb{T}^r_C = \mathbb{T}_L \mathbb{T}_{T_1} \mathbb{T}_\searrow \mathbb{T}_{T_1} \mathbb{T}_\nearrow$. (**b**) Without relaxation, $\mathbb{T}^{nr}_C = \mathbb{T}_L \mathbb{T}_\searrow \mathbb{T}_\nearrow$. The initial condition is given by the vector $\left(\frac{1}{4},\frac{1}{4},\frac{1}{4},\frac{1}{4},0,0,0,0\right)$.

We can now analyze further the effect of $T_1$ relaxation by addressing the case where $\beta_{up} \ll \beta_{down}$. More precisely, we consider the case $\beta_{up} \lesssim 5$ mT/ms and $\beta_{down}$ to be considerably faster. In practice, this translates into the following two assumptions:

1. $\beta_{up}$ is assumed to be *sufficiently slow* in order to ensure that, consistent with the SD limit, $p^\nearrow_{0,SD} \sim p^\nearrow_{0,\perp,SD} \sim 1/2$.
2. $\beta_{down}$ is assumed to be *sufficiently fast* in order to ensure that $p^\searrow_1 \sim 1$ and $p^\searrow_{1\perp} \sim 0$.

Under these assumptions, the two TMs for the forward and backward sweeps are:

$$\mathbb{T}_\nearrow = \begin{array}{c|cccccccc} & |1\rangle & |2\rangle & |3\rangle & |4\rangle & |5\rangle & |6\rangle & |7\rangle & |8\rangle \\ \hline \langle 1| & 1 & 0 & 0 & 0 & 0 & 0 & 0 & 0 \\ \langle 2| & 0 & 1 & 0 & 0 & 0 & 0 & 0 & 0 \\ \langle 3| & 0 & 0 & \frac{p_1^\nearrow}{2} & p_1^\nearrow p_{1\perp}^\nearrow & \frac{1}{2}(p_1^{\nearrow 2} + p_{1\perp}^{\nearrow 2}) & \frac{p_{1\perp}^\nearrow}{2} & 0 & 0 \\ \langle 4| & 0 & 0 & 0 & \frac{p_1^\nearrow}{2} & \frac{p_{1\perp}^\nearrow}{2} & \frac{1}{2} & 0 & 0 \\ \langle 5| & 0 & 0 & \frac{1}{2} & \frac{p_{1\perp}^\nearrow}{2} & \frac{p_1^\nearrow}{2} & 0 & 0 & 0 \\ \langle 6| & 0 & 0 & \frac{p_{1\perp}^\nearrow}{2} & \frac{1}{2}(p_1^{\nearrow 2} + p_{1\perp}^{\nearrow 2}) & p_1^\nearrow p_{1\perp}^\nearrow & \frac{p_1^\nearrow}{2} & 0 & 0 \\ \langle 7| & 0 & 0 & 0 & 0 & 0 & 0 & 1 & 0 \\ \langle 8| & 0 & 0 & 0 & 0 & 0 & 0 & 0 & 1 \end{array}$$

and

$$\mathbb{T}_\searrow = \begin{array}{c|cccccccc} & |1\rangle & |2\rangle & |3\rangle & |4\rangle & |5\rangle & |6\rangle & |7\rangle & |8\rangle \\ \hline \langle 1| & 1 & 0 & 0 & 0 & 0 & 0 & 0 & 0 \\ \langle 2| & 0 & 1 & 0 & 0 & 0 & 0 & 0 & 0 \\ \langle 3| & 0 & 0 & p_0^\searrow & 0 & p_{0\perp}^\searrow & 0 & 0 & 0 \\ \langle 4| & 0 & 0 & 0 & p_0^\searrow & 0 & p_{0\perp}^\searrow & 0 & 0 \\ \langle 5| & 0 & 0 & p_{0\perp}^\searrow & 0 & p_0^\searrow & 0 & 0 & 0 \\ \langle 6| & 0 & 0 & 0 & p_{0\perp}^\searrow & 0 & p_0^\searrow & 0 & 0 \\ \langle 7| & 0 & 0 & 0 & 0 & 0 & 0 & 1 & 0 \\ \langle 8| & 0 & 0 & 0 & 0 & 0 & 0 & 0 & 1 \end{array}$$

respectively. The cycle TM is then:

$$\mathbb{T}_C^r = \begin{array}{c|cccccccc} & |1\rangle & |2\rangle & |3\rangle & |4\rangle & |5\rangle & |6\rangle & |7\rangle & |8\rangle \\ \hline \langle 1| & 1/2 & 0 & \frac{(p_1^\nearrow + 1)}{4} & \frac{p_1^\nearrow p_{1\perp}^\nearrow}{2} + \frac{p_{1\perp}^\nearrow}{4} & \frac{p_1^{\nearrow 2}}{2} + \frac{p_{1\perp}^\nearrow}{4} & \frac{p_{1\perp}^\nearrow}{4} & 1/2 & 0 \\ \langle 2| & 0 & 1/2 & \frac{p_{1\perp}^\nearrow}{4} & \frac{p_1^{\nearrow 2}}{2} + \frac{p_{1\perp}^\nearrow}{4} & \frac{p_1^\nearrow p_{1\perp}^\nearrow}{2} + \frac{p_{1\perp}^\nearrow}{4} & \frac{(p_1^\nearrow + 1)}{4} & 0 & 1/2 \\ \langle 3| & 1/2 & 0 & \frac{(p_1^\nearrow + 1)}{4} & \frac{p_1^\nearrow p_{1\perp}^\nearrow}{2} + \frac{p_{1\perp}^\nearrow}{4} & \frac{p_1^{\nearrow 2}}{2} + \frac{p_{1\perp}^\nearrow}{4} & \frac{p_{1\perp}^\nearrow}{4} & 1/2 & 0 \\ \langle 4| & 0 & 1/2 & \frac{p_{1\perp}^\nearrow}{4} & \frac{p_1^{\nearrow 2}}{2} + \frac{p_{1\perp}^\nearrow}{4} & \frac{p_1^\nearrow p_{1\perp}^\nearrow}{2} + \frac{p_{1\perp}^\nearrow}{4} & \frac{(p_1^\nearrow + 1)}{4} & 0 & 1/2 \\ \langle 5| & 0 & 0 & 0 & 0 & 0 & 0 & 0 & 0 \\ \langle 6| & 0 & 0 & 0 & 0 & 0 & 0 & 0 & 0 \\ \langle 7| & 0 & 0 & 0 & 0 & 0 & 0 & 0 & 0 \\ \langle 8| & 0 & 0 & 0 & 0 & 0 & 0 & 0 & 0 \end{array}$$

(S.32)

It is straightforward to see that $\mathbb{T}_C^r$ is entirely ruled by $p_1^\nearrow$ and is independent of $p_0^\searrow$. This means that $\beta_{down}$ can be arbitrarily fast, as shown in Fig. 3(c) – main text. Additionally, the population imbalance generated after a cycle between states $|1\rangle$ and $|2\rangle$ is the same as the population imbalance between states $|3\rangle$ and $|4\rangle$. This is equivalent to the numerical observation made in Sec. IV, i.e. the generation of proton polarization becomes independent of the spin state of the P1 center.

The case with no $T_1$ relaxation corresponds to

$$\mathbb{T}_C^{nr} = \begin{array}{c|cccccccc} & |1\rangle & |2\rangle & |3\rangle & |4\rangle & |5\rangle & |6\rangle & |7\rangle & |8\rangle \\ \hline \langle 1| & 1 & 0 & \frac{p_0^{\searrow} + p_{0\perp}^{\searrow} p_1^{\nearrow}}{2} & p_{1\perp}^{\nearrow}\left(\frac{p_0^{\searrow}}{2} + p_{0\perp}^{\searrow} p_1^{\nearrow}\right) & \frac{p_0^{\searrow} p_1^{\nearrow}}{2} + \frac{p_{0\perp}^{\searrow}}{2}(p_1^{\nearrow 2} + p_{1\perp}^{\nearrow 2}) & \frac{p_{0\perp}^{\searrow} p_{1\perp}^{\nearrow}}{2} & 0 & 0 \\ \langle 2| & 0 & 1 & \frac{p_0^{\searrow} p_{1\perp}^{\nearrow}}{2} & \frac{p_{0\perp}^{\searrow} p_1^{\nearrow}}{2} + \frac{p_0^{\searrow}}{2}(p_1^{\nearrow 2} + p_{1\perp}^{\nearrow 2}) & p_{1\perp}^{\nearrow}\left(\frac{p_{0\perp}^{\searrow}}{2} + p_0^{\searrow} p_1^{\nearrow}\right) & \frac{1 - p_0^{\searrow} p_{1\perp}^{\nearrow}}{2} & 0 & 0 \\ \langle 3| & 0 & 0 & \frac{1 - p_0^{\searrow} p_{1\perp}^{\nearrow}}{2} & p_{1\perp}^{\nearrow}\left(\frac{p_{0\perp}^{\searrow}}{2} + p_0^{\searrow} p_1^{\nearrow}\right) & \frac{p_{0\perp}^{\searrow} p_1^{\nearrow}}{2} + \frac{p_0^{\searrow}}{2}(p_1^{\nearrow 2} + p_{1\perp}^{\nearrow 2}) & \frac{p_0^{\searrow} p_{1\perp}^{\nearrow}}{2} & 1 & 0 \\ \langle 4| & 0 & 0 & \frac{p_{0\perp}^{\searrow} p_{1\perp}^{\nearrow}}{2} & \frac{p_0^{\searrow} p_1^{\nearrow}}{2} + \frac{p_{0\perp}^{\searrow}}{2}(p_1^{\nearrow 2} + p_{1\perp}^{\nearrow 2}) & p_{1\perp}^{\nearrow}\left(\frac{p_0^{\searrow}}{2} + p_{0\perp}^{\searrow} p_1^{\nearrow}\right) & \frac{p_0^{\searrow} + p_{0\perp}^{\searrow} p_1^{\nearrow}}{2} & 0 & 1 \\ \langle 5| & 0 & 0 & 0 & 0 & 0 & 0 & 0 & 0 \\ \langle 6| & 0 & 0 & 0 & 0 & 0 & 0 & 0 & 0 \\ \langle 7| & 0 & 0 & 0 & 0 & 0 & 0 & 0 & 0 \\ \langle 8| & 0 & 0 & 0 & 0 & 0 & 0 & 0 & 0 \end{array}$$

The expression above shows an explicit dependence on $p_0^{\searrow}$. In the limit case of very fast $\beta_{down}$, we have $p_0^{\searrow} \to 1$ and $p_{0\perp}^{\searrow} \to 0$. Then,

$$\mathbb{T}_C^{nr} = \begin{array}{c|cccccccc} & |1\rangle & |2\rangle & |3\rangle & |4\rangle & |5\rangle & |6\rangle & |7\rangle & |8\rangle \\ \hline \langle 1| & 1 & 0 & \frac{1}{2} & \frac{p_{1\perp}^{\nearrow}}{2} & \frac{p_1^{\nearrow}}{2} & 0 & 0 & 0 \\ \langle 2| & 0 & 1 & \frac{p_{1\perp}^{\nearrow}}{2} & \frac{1}{2}(p_1^{\nearrow 2} + p_{1\perp}^{\nearrow 2}) & p_{1\perp}^{\nearrow} p_1^{\nearrow} & \frac{p_1^{\nearrow}}{2} & 0 & 0 \\ \langle 3| & 0 & 0 & \frac{p_1^{\nearrow}}{2} & p_{1\perp}^{\nearrow} p_1^{\nearrow} & \frac{1}{2}(p_1^{\nearrow 2} + p_{1\perp}^{\nearrow 2}) & \frac{p_{1\perp}^{\nearrow}}{2} & 1 & 0 \\ \langle 4| & 0 & 0 & 0 & \frac{p_1^{\nearrow}}{2} & \frac{p_{1\perp}^{\nearrow}}{2} & \frac{1}{2} & 0 & 1 \\ \langle 5| & 0 & 0 & 0 & 0 & 0 & 0 & 0 & 0 \\ \langle 6| & 0 & 0 & 0 & 0 & 0 & 0 & 0 & 0 \\ \langle 7| & 0 & 0 & 0 & 0 & 0 & 0 & 0 & 0 \\ \langle 8| & 0 & 0 & 0 & 0 & 0 & 0 & 0 & 0 \end{array} \quad (S.33)$$

A comparison between the $\mathbb{T}_C^r$ and $\mathbb{T}_C^{nr}$, as given by Eqns. (S.32) and (S.33) respectively, is shown in Fig. S7. There, we show the proton polarization as a function of both the number of cycles and $p_1^{\nearrow}$. While a finite $T_1$ time ensures proton polarization close to 50%, the case without relaxation can go beyond 10% only by substantially decreasing $p_1^{\nearrow}$. But even that situation is not possible in practice, because other low-velocity effects (namely, dephasing or spin-lattice relaxation during the LZ crossings) would compromise efficiency.

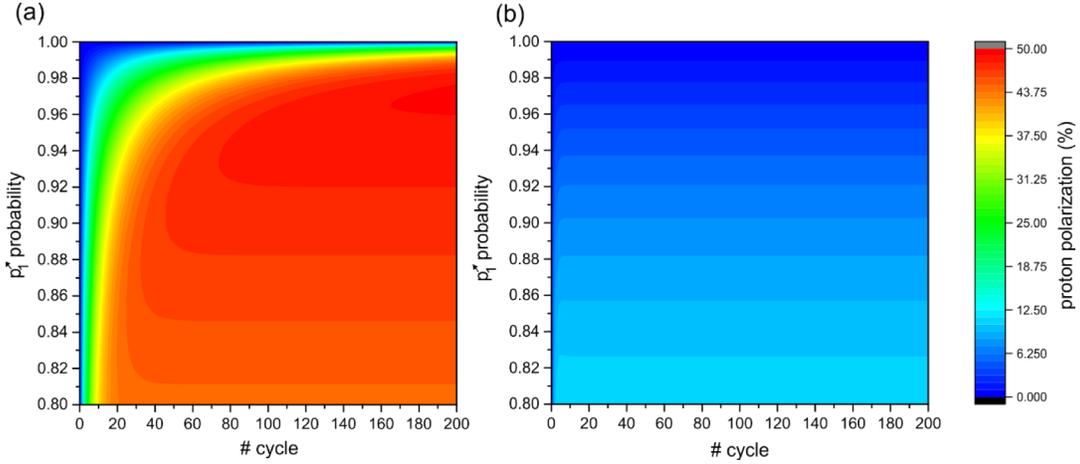

**Figure S7.** Proton polarization evaluated using the TM approach, as a function of the number of cycles and $p_1^\nearrow$. (**a**) The cycle TM is given by Eq. (S.32), i.e. $T_1$ relaxation is considered. (**b**) The cycle TM is given by Eq. (S.33), i.e. $T_1$ relaxation is not considered. In both cases the initial condition is given by the vector $\left(\frac{1}{4},\frac{1}{4},\frac{1}{4},\frac{1}{4},0,0,0,0\right)$.

## S.VII The impact of other paramagnetic impurities

In a realistic scenario, the NV center could simultaneously interact with some other P1 centers not necessarily close to the surface (i.e. not coupled to any proton); such a number ultimately depends on the implantation density. In practice, the question that arises is whether the presence of at least one 'bystander' P1 center could destroy the efficiency in the polarization transfer. Addressing this problem in full is a difficult task, as reproducing the dynamics of a many-spin system is a computationally challenging problem beyond the scope of this work. Instead, we consider one of the worst possible scenarios where the NV — weakly coupled to a 'proxy' P1 center mediating the polarization transfer to the 'target' $^1$H — interacts strongly with a 'bystander' P1 in close proximity.

Fig. S8(a) shows a schematic of the spin geometry: In our calculations, we set $J_{\text{dip}}^{\text{NV}-\text{P1}} = 300$ kHz, and $J_{\text{dip}}^{\text{H}-\text{P1}} = 200$ kHz; on the other hand, the 'bystander' P1 — denoted here as 'B1' — is set to couple to the NV via $J_{\text{dip}}^{\text{NV}-\text{B1}} = 1$ MHz (but does not interact with the proxy or target spins). Near 51 mT, the energy diagram now displays a considerable more complex structure (compare with Figs. S4 and S5) with two sets of level anti-crossings. The first set, shown in Fig. S8(b), mixes states $\left|0,+\frac{1}{2},m_I,-\frac{1}{2}\right\rangle$ and $\left|0,-\frac{1}{2},m_I,+\frac{1}{2}\right\rangle$ with states of the form $\left|-1,-\frac{1}{2},m_I,-\frac{1}{2}\right\rangle$ (all labels indicate spin quantum projection numbers in the order $|\text{NV}, \text{P1}, {^1}\text{H}, \text{B1}\rangle$). The second set, shown in Fig. S8(c), brings together states $\left|0,+\frac{1}{2},m_I,+\frac{1}{2}\right\rangle$, $\left|-1,-\frac{1}{2},m_I,+\frac{1}{2}\right\rangle$, and $\left|-1,+\frac{1}{2},m_I,-\frac{1}{2}\right\rangle$.

To benchmark the efficiency of the protocol we first compare in Fig. S8(d) the proton polarization upon a single low-to-high sweep at $\beta = 0.25$ mT/ms for spin clusters with and without a bystander spin. In both cases we start with the NV polarized in $|0\rangle$ and assume a thermal state for all other spins. We find that even if partly reduced, the end nuclear spin polarization is still quite substantial. The reason is that when starting from states $|1\rangle = \left|0,-\frac{1}{2},\uparrow,+\frac{1}{2}\right\rangle$ and $|3\rangle = \left|0,-\frac{1}{2},\downarrow,+\frac{1}{2}\right\rangle$

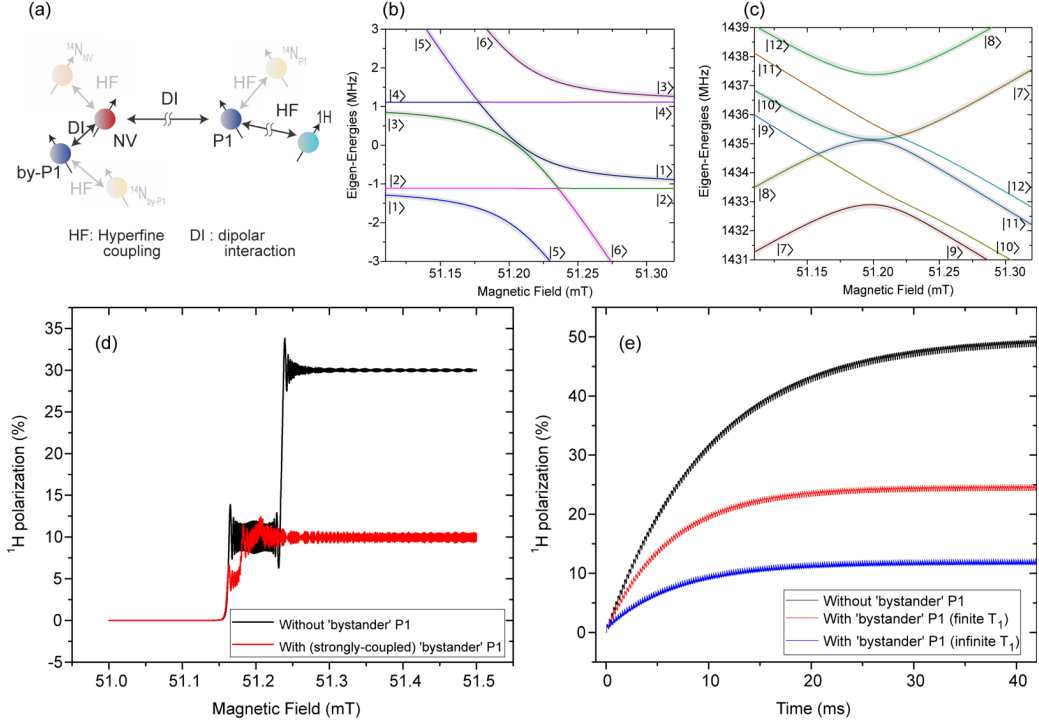

**Figure S8**. Spin dynamics in the presence of a strongly-coupled 'bystander' P1. (**a**) Schematics of the spin cluster. Here, $J_{dip}^{NV-P1} = 300$ kHz, $J_{dip}^{H-P1} = 200$ kHz, and $J_{dip}^{NV-B1} = 1$ MHz. We assume the magnetic field to be aligned with the NV axis, i.e. $\theta = 0$, $\phi = 0$. We ignore the NV and P1 hyperfine interactions with their $^{14}$N hosts (faded spins). (**b**) Eigen-energies close to the resonance, first set of avoided crossings. We denote $|1\rangle = \left|0, -\frac{1}{2}, \uparrow, +\frac{1}{2}\right\rangle$, $|2\rangle = \left|0, +\frac{1}{2}, \uparrow, -\frac{1}{2}\right\rangle$, $|3\rangle = \left|0, -\frac{1}{2}, \downarrow, +\frac{1}{2}\right\rangle$, $|4\rangle = \left|0, +\frac{1}{2}, \downarrow, -\frac{1}{2}\right\rangle$, $|5\rangle = \left|-1, -\frac{1}{2}, \uparrow, -\frac{1}{2}\right\rangle$, and $|6\rangle = \left|-1, -\frac{1}{2}, \downarrow, -\frac{1}{2}\right\rangle$, see text for the label order. (**c**) Eigen-energies close to the resonance for the second set of avoided crossings. We use the notation $|7\rangle = \left|0, +\frac{1}{2}, \uparrow, +\frac{1}{2}\right\rangle$, $|8\rangle = \left|0, +\frac{1}{2}, \downarrow, +\frac{1}{2}\right\rangle$, $|9\rangle = \left|-1, -\frac{1}{2}, \uparrow, +\frac{1}{2}\right\rangle$, $|10\rangle = \left|-1, +\frac{1}{2}, \uparrow, -\frac{1}{2}\right\rangle$, $|11\rangle = \left|-1, -\frac{1}{2}, \downarrow, +\frac{1}{2}\right\rangle$, and $|12\rangle = \left|-1, +\frac{1}{2}, \downarrow, -\frac{1}{2}\right\rangle$. The faint shading on some of the traces in (b) and (c) highlights the subset of levels with effective topology equivalent to that in Figs. S4 and S5. (**d**) Single low-to-high sweep at $\beta = 0.25$ mT/ms. (**e**) Multiple cycles with $\beta_{up} = 6$ mT/ms and $\beta_{down} = 10$ mT/ms. The blue trace corresponds to the case where all P1s have 'infinite' spin-lattice relaxation times, whereas the red trace displays the results for $\tau_{LZ} < T_1^{(P1)} \lesssim T_c/2$.

in Fig. S8(b) (or states $|7\rangle = \left|0, +\frac{1}{2}, \uparrow, +\frac{1}{2}\right\rangle$, and $|8\rangle = \left|0, +\frac{1}{2}, \downarrow, +\frac{1}{2}\right\rangle$ in Fig. S8(c)), the effective topology of the energy diagram can be made analogous to the bystander-free diagrams of Figs. S4 and S5, provided the field sweep rate is fast enough to cancel the effect of small gaps (shaded traces in Figs. S8(b) and S8(c)). In Fig. S8(e) we go one step farther and compare the polarization buildup upon multiple field cycles using $\beta_{up} = 3$ mT/ms and $\beta_{down} = 10$ mT/ms. As in the previous case, we observe significant DNP, particularly when both the proxy and bystander P1s undergo rapid spin relaxation, an observation consistent with the findings discussed in the main text (Fig. 3).

## S.VIII Summary features and operating range

As a quick reference to the findings in this work, Table S1 shows a summary of the key aspects of our protocol as well as the parameter regime where this approach is expected to yield significant DNP.

Table S1.

| Summary features and operating ranges |
|---|
| • Positive (negative) nuclear polarization produced for NV optical pumping at the low (high) end of the sweep. |
| • Nuclear polarization attained for field sweep rates in the range 0.5-10 mT/ms, though the use of faster rates is possible when $\tau_{LZ} < T_1^{(P1)} \lesssim T_c/2$. Efficient DNP is expected even in the case where $T_2^{(P1)} < \tau_{LZ}$. |
| • Continuous illumination yields nuclear polarization provided one half of the magnetic field cycle is faster than the other. |
| • The polarization transfer is robust to NV-field misalignment up to θ∼25 deg. |
| • Operates over a broad range of NV–P1 coupling conditions, down to 300 kHz (corresponding to a ~5.4 nm NV–P1 distance). |
| • Efficient polarization transfer to target nuclei is expected for P1–$^1$H couplings down to ~200 kHz (corresponding to P1–$^1$H distances of ~0.8 nm). Note that proton spins need not be coupled to the NV. |
| • DNP robust against the presence of P1s other than that coupled to the target $^1$H; efficient polarization transfer expected for P1 concentrations of up to ~50 ppm. |